\documentclass[a4paper,11pt]{article}
\pdfoutput=1 

\usepackage{jheppub} 
\usepackage[T1]{fontenc} 
\usepackage[italian,english]{babel}
\usepackage{hyperref}
\usepackage{ifpdf}
\usepackage{subfigure}
\usepackage{tikz}
\usepackage[compat=1.1.0]{tikz-feynman}
\usepackage{slashed}
\usetikzlibrary{arrows,shapes}
\usetikzlibrary{trees}
\usetikzlibrary{matrix,arrows} 
\usetikzlibrary{positioning}
\usetikzlibrary{calc,through}
\usetikzlibrary{decorations.pathreplacing}  
\usepackage{pgffor}

\usetikzlibrary{decorations.pathmorphing}
\usetikzlibrary{decorations.markings}
\tikzset{
vector/.style={decorate, decoration={snake}, draw},
fermion/.style={draw=black, postaction={decorate}}, 
scalar/.style={dashed,draw=black, postaction={decorate}}}
\tikzstyle{block} = [draw, rectangle, 
minimum height=3em, minimum width=6em]

\usepackage{amssymb}
\usepackage{amsfonts}
\usepackage{epsf}
\usepackage{rotating}
\usepackage{graphicx}
\usepackage{amsmath}
\usepackage{fancyhdr}
\usepackage{lineno}
\usepackage{babel}
\usepackage{graphics}
\usepackage{pstricks}
\usepackage{color}
\usepackage{multirow}
\usepackage{float}
\usepackage{lineno}
\usepackage{mathtools}
\usepackage{stackengine}
\usepackage{comment}

\newcommand{\lsim}{\mathrel{\mathop{\kern 0pt \rlap
{\raise.2ex\hbox{$<$}}}
\lower.9ex\hbox{\kern-.190em $\sim$}}}
\newcommand{\gsim}{\mathrel{\mathop{\kern 0pt \rlap
{\raise.2ex\hbox{$>$}}}
\lower.9ex\hbox{\kern-.190em $\sim$}}}

\newcommand{\be}{\begin{equation}}
\newcommand{\ee}{\end{equation}}
\newcommand{\bea}{\begin{eqnarray}}
\newcommand{\eea}{\end{eqnarray}}

\def\gev{\ensuremath{\mathrm{\,Ge\kern -0.1em V\,}}}
\def\tev{\ensuremath{\mathrm{\,Te\kern -0.1em V\,}}}

\begin{document}

\rightline{DESY-26-065}

\title{ Pseudo-Nambu-Goldstone inflation with $Z_N$ symmetric waterfall fields }

\author[a]{Hyun Min Lee}
\author[b]{and Adriana G. Menkara}

\affiliation[a]{Department of Physics, Chung-Ang University, Seoul 06974, Korea.}
\affiliation[b]{Deutsches Elektronen-Synchrotron DESY, Notkestr. 85, 22607 Hamburg, Germany.}

\emailAdd{hminlee@cau.ac.kr}
\emailAdd{adriana.menkara@desy.de}


\abstract{ We propose a hybrid inflation model where a pseudo-Nambu-Goldstone boson inflaton couples to $N$ waterfall scalar fields respecting a $Z_N$ symmetry. We identify the phases for the inflation and the consequent waterfall transition, concretely, in $Z_2$, $Z_3$ and $Z_4$ cases. From the Coleman-Weinberg potential for the inflaton, we show that  the quadratically divergent corrections coming from the waterfall sector are cancelled due to the $Z_N$ symmetry, while the logarithmically divergent corrections are absent only for $N>2$, ensuring the radiative stability of the inflaton potential. We show the parameter space for a successful inflation with the loop-corrected inflaton potential in each model and compare the results between different discrete symmetries. We further analyze the vacuum structure of the models and the reheating process due to the $Z_N$-invariant Higgs-portal couplings for the waterfall fields. We find that the maximum temperature can be smaller than the mass of the waterfall  field condensate such that the $Z_N$ symmetry is not restored after reheating and there is no domain wall problem in the models. We also comment on the possibility of multi-component dark matter from the $Z_N$ partners of the waterfall field condensate.
}

\maketitle              

\section{Introduction}

Cosmic inflation has been regarded as the paradigm for standard cosmology just after Big Bang, but we have no clue for either the origin of the inflaton potential or the recovery of the radiation-dominated universe after inflation (the so called reheating). Measurements of anisotropies of Cosmic Microwave Background (CMB) \cite{Planck} not only give rise to a precise determination of the cosmological parameters, together with high-$z$ supernova data and galaxy surveys at large scales, but also favor a slow-roll inflation with a single scalar field via the spectral index and the $B$-mode polarizations of the CMB photons \cite{keck}. 

There are still plenty of possibilities for the inflaton potential that are consistent with the current CMB measurements, although some of simple monomial potentials and a certain class of natural inflation models have been already excluded by the null observation of the primordial $B$-mode polarizations \cite{Planck,keck}.  Recently, the measurements of ACT at small scales combined with large-scale structures \cite{ACT} show an increase of the spectral index towards scale-invariance as compared to Planck alone, so the loop corrections in the well-fitted inflation models with Planck have been included in light of the ACT results \cite{Han:2025cwk}.  Therefore, it would be important to pin down a more precise value of the spectral index and measure the primordial $B$-mode polarizations in the future CMB experiments.

Hybrid inflation \cite{hybrid} is an alternative possibility to single-field inflation where inflation ends due to the transition of the waterfall scalar field even if the inflaton does not violate the slow-roll conditions. In this case, it is necessary to introduce a sizable coupling of the waterfall field to the inflaton in order for  the waterfall field to develop a tachyonic instability as the inflaton rolls down the potential. The very existence of the waterfall field coupling to the inflaton could destroy the flatness of the inflaton potential by loop corrections, so a small inflaton mass as compared to the waterfall field mass and the cutoff scale might call for a natural explanation.

In this regard, a natural candidate for the inflaton is a pseudo-Nambu-Goldstone(pNGB) boson \cite{natural,multiaxion1,multiaxion2}, originating from a spontaneously broken global symmetry, whose mass is protected by the shift symmetry at the perturbative level, although it is broken explicitly by non-perturbative effects with anomalies, and it could be subject to quantum gravity effects \cite{qgravity}. On the other hand, the quadratically divergent  loop corrections to the inflaton potential are prohibited in the presence of the $Z_2$ symmetric couplings of the twin waterfall fields to the inflaton \cite{z2,Lee:2022fkd,Lee:2023dcy}, being  the union for hybrid inflation  of the relaxion \cite{relaxion} and the twin symmetry \cite{twin}.

In this article, we propose a $Z_N$ symmetric extension of the pNGB inflation with $N$ waterfall scalar fields. In this model,  both the inflaton potential and the couplings of the waterfall fields to the inflaton respect a $Z_N$ symmetry, as for the solution to the little hierarchy problem for the Higgs mass \cite{hook}. It is remarkable that such discrete symmetries, if originated from gauge symmetries, could survive quantum gravity effects \cite{discrete}. 
We regard all the waterfall fields as being stabilized at the origin during inflation and the inflation ends when one of the waterfall fields undergoes a waterfall transition. We identify the Configurations for the inflation vacua and the waterfall transitions, depending on the order of the discrete symmetry, for instance, $Z_2, Z_3$, and $Z_4$ symmetries.  

We study the role of the $Z_N$ symmetry in keeping the inflaton potential flat during inflation against the loop corrections of the waterfall fields. In particular, we show how the logarithmic loop corrections to the inflaton potential change the parameter space for a successful inflation in the $Z_2$ case. We also discuss the case with  higher order $Z_N$ symmetries with $N>2$ and compare it with the $Z_2$ case. We identify the vacuum structure for the inflaton and the waterfall sector and consider the reheating from the decays of the waterfall fields via  the $Z_N$-invariant Higgs-portal couplings. If there are accidental $Z'_2$ symmetries for the  waterfall fields that are orthogonal to the direction of the waterfall transition,  we consider the possibility of multi-component dark matter from those $Z_N$ partners.

\section{The setup}

We introduce a pseudo-Nambu-Goldstone boson $\phi$ as the inflaton and $N$ real scalar fields $\chi_k (k=1,2,\cdots, N)$ as waterfall fields. 
Then, imposing the $Z_N$ symmetry under which $\chi_k\to \chi_{k+1}$ and $\phi\to \phi+ 2\pi f/N$, we introduce the general $Z_N$ invariant potential for the inflaton and the waterfall fields with nearest neighbor interactions only \footnote{We note that the mixing mass terms  with $\alpha$ can be zero due to $Z'_2$ symmetries or separate symmetries in the mirror sectors. For instance, for $N={\rm even}$, the $Z'_2$ parities are $\chi_k\to \chi_k$ and $\chi_{k+1}\to -\chi_{k+1}$ for $ k=1,2,\cdots, N$.  Then, some of the waterfall fields with zero VEVs can be dark matter candidates  \cite{Lee:2022fkd,Lee:2023dcy}. Even if the mixing mass terms are nonzero, it was shown in the $Z_2$ case  that the mass mixing terms do not affect the cancellation of quadratic divergent corrections to the Coleman-Weinberg potential during inflation \cite{Lee:2022fkd,Lee:2023dcy}, so similar conclusions can be drawn for the $Z_N$ cases. The effects of $\alpha\neq 0$ will be also discussed for the $Z_N$ case later. }, as follows,
\bea
V(\phi,\chi_k)=V_{\rm inf}(\phi)+V_{W}(\phi,\chi_k), \label{fullpot}
\eea
with
{\small
\bea
V_{\rm inf}(\phi)&=& V_0+ \Lambda^4 \cos\Big( \frac{N\phi}{f}+\delta\Big) , \\
V_W(\phi,\chi_k)&=& \sum_{k=1}^N \bigg[\frac{1}{2}m^2_\chi  \chi^2_k -\frac{1}{2}\mu^2 \chi^2_k \cos\Big(\frac{\phi}{f}+ \frac{2\pi k}{N} \Big) +\alpha^2 \chi_k \chi_{k+1}+\frac{1}{4} \lambda \chi^4_k+\frac{1}{2}\lambda'  \chi^2_k \chi^2_{k+1}\bigg], 
 \label{treepot}
\eea
}
where $V_0$ is a constant vacuum energy, $\delta$ is a constant phase shift,  $\chi_{N+1}=\chi_1$ and $\chi_{N+2}=\chi_2$.
Here, we didn't consider the interactions between non-consecutive  waterfall fields such as $\sum_k \chi_k \chi_{k+2}$, etc. This could be justified in the case when the waterfall fields are located at distinct sites in theory space or in the extra dimension as in the clockwork models so they are connected only by the nearest neighbor interactions \cite{clockwork}. We also note that other oscillating potentials, $\cos(lN\phi/f+\delta_l)$, with $l>1$, are consistent with the $Z_N$ symmetry, but it is originated from a higher $Z_{M}$ discrete symmetry with $M=lN$ in the UV. In our work, assuming that there is no higher order discrete symmetry surviving in the IR, we focus on the minimal case with the $Z_N $ symmetry without such UV potentials.

We also assume that a $Z_2$ symmetry making all the waterfall fields transforming by $\chi_k\to-\chi_k$ simultaneously,  or global $U(1)$ symmetries on the waterfall fields, are imposed to forbid all the interaction terms containing an odd number of the waterfall fields. Moreover, in the presence of separate $Z_2$ symmetries, namely, $\chi_k\to -\chi_k$, with $k=1,2,\cdots, N$, in which case $
\alpha=0$, some of the waterfall fields can be dark matter candidates, as will be discussed later. But, when the interaction terms contain only the even number of waterfall fields, there are potential problems with stable domain-walls for a high reheating temperature, so bias terms violating the extra symmetries can be introduced to avoid the domain-wall problems.

As a result, the effective mass parameters for the waterfall fields are
\bea
m^2_{\chi,k} = m^2_\chi -\mu^2 \cos\Big(\frac{\phi}{f}+ \frac{2\pi k}{N} \Big)\equiv m^2_\chi+m^2_k,\quad k=1,2,\cdots, N.
\eea
Henceforth, we assume that $m^2_\chi, \mu^2>0$ and $\mu^2>m^2_\chi$ for the waterfall transition. We note that the quartic couplings of the waterfall fields are subject to the vacuum stability and perturbative unitarity bounds, as follows,
\bea
&&\lambda>0, \quad \lambda+\lambda'>0, \quad \lambda+2\lambda'>0, \,\, {\rm for} \,\, \lambda'<0, \\
&&\lambda <\frac{8\pi}{3}, \quad |\lambda'|< 8\pi.
\eea

\subsection{Tree-level effective potential for inflaton}

After integrating out the waterfall fields for $\lambda'=0$ and $\alpha=0$, the effective potential for $\phi$ at tree level is generated,  as follows,
\bea
V_{1,\rm eff}&=& -\frac{1}{4\lambda} \sum_{k=1}^N m^4_{\chi,k}\theta(-m^2_{\chi,k}).
\eea
As we will show in the examples for $N=2,3,4$ later, it is not possible that all scalar fields $\chi_k$'s acquire a non-vanishing VEV. Nonetheless, for the configuration where all the waterfall fields get nonzero VEVs, the above effective potential becomes
\bea
V_{1,\rm eff}&=& -\frac{1}{4\lambda} \sum_{k=1}^N (m^2_\chi+m^2_k)^2  \nonumber \\
&=&-\frac{N}{4\lambda}\, m^4_\chi -\frac{1}{2\lambda} m^2_\chi  \sum_{k=1}^N m^2_k  -\frac{1}{4\lambda} \sum_{k=1}^N m^4_k.  \label{effpot}
\eea
Using the following identities for the sums,
\bea
 \sum_{k=1}^N m^2_k &=& -\mu^2  \sum_{k=1}^N \cos\Big(\frac{\phi}{f}+\frac{2\pi k}{N} \Big)=0,  \,\,\, N\geq 2, \label{sum1} \\
 \sum_{k=1}^N m^4_k&=& \mu^4 \sum_{k=1}^N \cos^2\Big(\frac{\phi}{f}+\frac{2\pi k}{N} \Big)=\frac{N}{2}\, \mu^4, \,\,\, N>2, \label{sum2}
\eea
we would get a constant effective potential $V_{1,{\rm eff}}$. Then, the effective potential would be trivially $Z_N$ symmetric and the inflaton potential would be flat at tree level. Here, for the $N=2$ case,  the identity in eq.~(\ref{sum2}) is changed to $2\mu^4 \cos^2\big(\frac{\phi}{f}\big)$, so the effective potential would depend on $\phi$. But, it turns out that there is no vacuum satisfying all $m^2_{\chi,k}<0$ simultaneously, independently of $N$. This is in contrast  with Ref.~\cite{hook} where the $Z_N$ symmetries for the mirror sectors with $m^2_\chi, \mu^2<0$ and $|m^2_\chi|<|\mu^2|$ were introduced to  solve the little hierarchy problem for the Higgs mass so there exists a vacuum with all $m^2_{\chi,k}<0$.

For a natural hybrid inflation, we make a choice of the mass parameters for $m^2_\chi$ and  $\mu^2$, namely, $m^2_\chi, \mu^2>0$ and $m^2_\chi<\mu^2$. We then choose the initial value of the inflaton such that all the effective square masses of the waterfall scalar fields are positive, and inflation occurs. The effective masses for the waterfall fields change as the inflaton evolves along the resulting potential, until one of them becomes negative. When this happens, the waterfall transition occurs for the field whose effective square mass becomes negative, thus ending inflation. We show how this concretely occurs in the $N = 2,3,4$ examples that we study later.

For Configuration I (the inflation phase) where all the VEVs of all the waterfall field vanish,  there is no effective potential for the inflaton at tree level during inflation, namely, $V_{1,{\rm eff}}=0$. Configuration I allows for the initial condition for a natural hybrid inflation without quadratic sensitivity to the UV physics due to the $Z_N$ symmetry.

Now we also discuss Configuration II (the waterfall phase) where the VEVs of some $M$ waterfall fields with $M<N$ vanish, i.e., $\langle \chi_a\rangle=0$, with $a=1,2,\cdots, M$.
In this case, only the waterfall fields with nonzero VEVs contribute to the effective potential for $\phi$ at tree level, leading to 
\bea
V_{1,\rm eff}&=& -\frac{1}{4\lambda} \sum_{k\neq a} m^4_{\chi,k}\nonumber \\
&=& -\frac{1}{4\lambda}  \sum_{k=1}^N m^4_{\chi,k} + \frac{1}{4\lambda}\sum_{a=1}^M m^4_{\chi,a} \nonumber \\
&=&\frac{1}{4\lambda}\sum_{a=1}^M m^4_{\chi,a}  +{\rm constant} \nonumber \\
&=&\frac{1}{4\lambda}\sum_{a=1}^M \left[m^2_\chi-\mu^2\cos\Big(\frac{\phi}{f}+\frac{2\pi a}{N}\Big) \right]^2+{\rm constant}.  \label{effpot2}
\eea
Here, the third equality is true only for $N>2$, because the full sum in eq.~(\ref{sum2}) is constant for $N>2$.
In this case, the effective potential for $\phi$ becomes nonzero due to the missing contributions from the waterfall fields with nonzero VEVs.  
We note that if the inflaton is stabilized near the boundary between Configurations I and II,  the waterfall field with a nonzero VEV can become light during the transition from Configuration I to Configuration II.

\subsection{One-loop Coleman-Weinberg potential for inflaton}

Now we discuss the one-loop CW potential for the inflaton due to waterfall fields.

The one-loop CW potential for the inflaton due to  $N$ waterfall fields in the cutoff regularization, is given by
\bea
V_{\rm CW}= \frac{1 }{64\pi^2} \sum_{k=1}^N \bigg[2 M^2_{\chi,k}M^2_* -M^4_{\chi,k} \ln \frac{e^{\frac{1}{2}}M^2_*}{m^2_{\chi,k}}\bigg] \label{CW}
\eea
where  $M_*$ is the cutoff scale and $M^4_{\chi,k}$ are the effective masses for waterfall fields. 
For $\lambda'=0$ and $\alpha=0$, the effective masses for waterfall fields are given by $M^2_{\chi,k}=3\lambda \chi^2_k+m^2_{\chi,k}$. Consequently, $M^2_{\chi,k}\to M^2_{\chi,k+1}$ under the $Z_N$ symmetry, namely, $\chi_k\to \chi_{k+1}$ and $\phi\to \phi+2\pi f/N$, so the above CW potential (\ref{CW}) is invariant under the $Z_N$ symmetry.
Then, using the sum rule for the effective squared masses for the waterfall fields,  we obtain the CW potential as
\bea
V_{\rm CW}
=  \frac{1}{32\pi^2} \Big(N m^2_\chi  +3\lambda \sum_{k=1}^N\chi^2_k\Big) M^2_*- \frac{1}{64\pi^2} \sum_{k=1}^N  \Big(m^2_{\chi,k}+3\lambda\chi^2_k\Big)^2 \ln \frac{e^{\frac{1}{2}}M^2_*}{M^2_{\chi,k}}. \label{CW2}
\eea
In this case, the quadratically divergent terms for the inflaton are cancelled between the waterfall fields due to the sum rule in eq.~(\ref{sum1}) and there appears only logarithmic corrections for the inflaton potential, depending on the VEVs of the waterfall fields \cite{Lee:2022fkd,Lee:2023dcy}.  
We remark that when the VEVs of all the waterfall fields vanish as during inflation, we get even the logarithmically divergent terms for the inflaton cancelled for the $Z_N$ case with $N>2$, because of the sum rule in eq.~(\ref{sum2}), unlike the $Z_2$ case where the sum of the $m^4_k$ terms is not constant.

On the other hand, the waterfall fields receive quadratically divergent corrections to their masses proportional to the quartic coupling $\lambda$, being UV sensitive.
Nonetheless, we can protect the inflaton potential from getting huge corrections by the waterfall field couplings, thanks to the $Z_N$ symmetry.

\section{Phases for waterfall transitions}

For simplicity,  we consider the case with vanishing mixing masses for the waterfall fields, $\alpha=0$.
We also set the mixing  quartic couplings for the waterfall fields, $\lambda'=0$, but our qualitative results such as the classification of the inflation vacua do not depend on either the mixing masses or the mixing quartic couplings, as will be discussed later in the section and in the Appendix.

\subsection{The $Z_2$ model}

The waterfall effective mass parameters for $N=2$ \cite{Lee:2022fkd,Lee:2023dcy} are given by
\bea
m^2_{\chi,1} &=& m^2_\chi +\mu^2\cos\Big(\frac{\phi}{f}\Big), \label{z2m1} \\
m^2_{\chi,2} &=& m^2_\chi -\mu^2\cos\Big(\frac{\phi}{f}\Big). \label{z2m2}
\eea
In this case, taking $x\equiv \cos\Big(\frac{\phi}{f} \Big)/a_0$  with $a_0\equiv \frac{m^2_\chi}{\mu^2}<1$, Configuration I for $N=2$ is realized for $|x|<1$. This is because we have $m^2_{\chi,1}, m^2_{\chi,2}>0$ for $|x|<1$, maintaining the unbroken symmetries in the mirror sectors. But, as the field $\phi$ moves along during inflation from $|x|<1$ to $|x|>1$, $m^2_{\chi,2}$(or $m^2_{\chi,1}$) scans from positive to negative values, driving a waterfall transition.  

In Configuration II where  $\langle \chi_1\rangle= 0$ and  $\langle \chi_2\rangle\neq 0$, the effective potential at tree level  becomes 
\bea
V_{1,\rm eff}(\phi)&=& -\frac{1}{4\lambda} m^4_{\chi,2} +{\rm const}  \nonumber \\
&=&-\frac{\mu^4}{4\lambda}  \Big( \cos \Big(\frac{\phi}{f} \Big)-a_0\Big)^2 +{\rm constant}. 
\eea
Then, from the above effective potential, we get
\bea
V^\prime_{1,\rm eff}&=&\frac{\mu^4}{2f\lambda} \bigg(\cos \Big(\frac{\phi}{f} \Big)-a_0\bigg)\sin \Big(\frac{\phi}{f} \Big), \\
V^{\prime\prime} _{1,\rm eff}&=& \frac{\mu^4}{2f^2 \lambda}\,\bigg(2\cos^2 \Big(\frac{\phi}{f} \Big)-1-a_0 \cos \Big(\frac{\phi}{f} \Big)  \bigg).
\eea
Thus, after $m^2_{\chi,2}$ changes its sign from positive to negative, there is no stable minimum of $V_{1,\rm eff}$ with $V^{\prime\prime}_{1,{\rm eff}}>0$ in Configuration II, but we can stabilize the Configuration II by the inflaton potential ($V_{\rm inf}$) in the true vacuum.

We note that there is no configuration where two waterfall fields have $m^2_{\chi_1},m^2_{\chi_2} <0$ so their VEVs cannot be nonzero simultaneously \footnote{ A mixing mass $\chi_1\chi_2$ and a  mixing quartic coupling, $\chi^2_1\chi^2_2$, that are consistent with the $Z_2$ symmetry, was introduced in Refs.~\cite{Lee:2022fkd,Lee:2023dcy}, so there can exist a vacuum with $\langle \chi_1\rangle\neq 0$ and $\langle \chi_2\rangle\neq 0$. }.

\subsection{The $Z_3$ model}

\begin{figure}[!t]
\begin{center}
\includegraphics[width=0.45\textwidth]{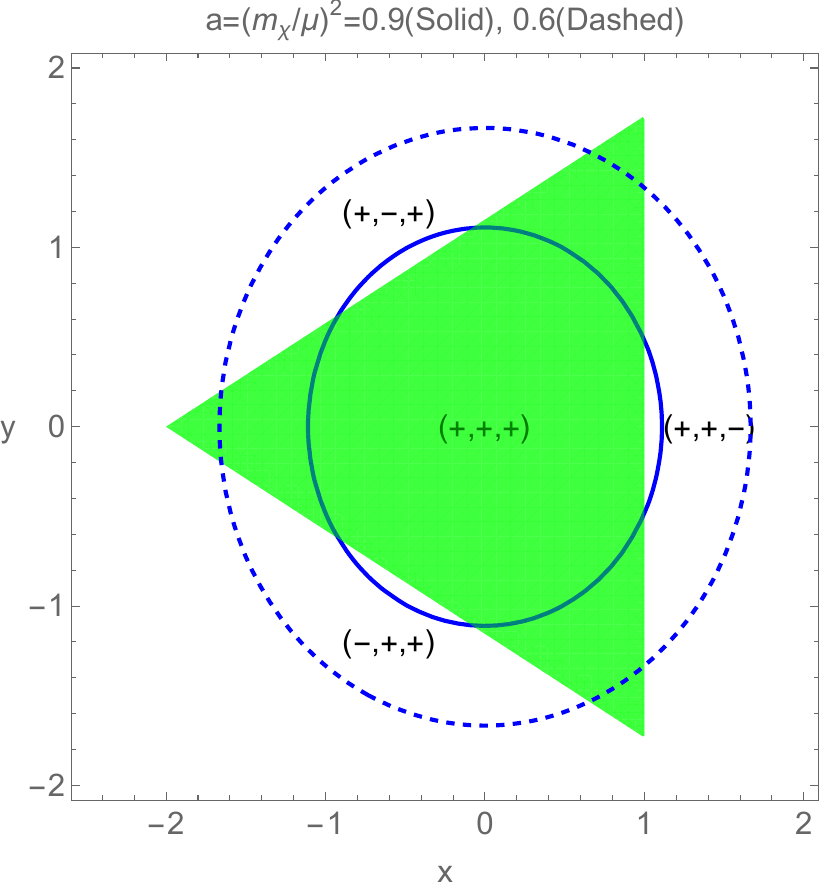}
\caption{Phase of the field $\phi$ in $x=\cos(\phi/f)/a_0$ versus $y=\sin(\phi/f)/a_0$ for the $Z_3$ case. We chose $a_0\equiv \frac{m^2_\chi}{\mu^2}=0.9, 0.6$ in solid and dashed lines, respectively. Configuration I (the inflation phase) is realized only along the part of solid or dashed lines inside the green triangle. We have also indicated (${\rm sgn}(m^2_{\chi_1}),{\rm sgn}(m^2_{\chi_2}),{\rm sgn}(m^2_{\chi_3}))$ in Configuration II (the waterfall phase) connected to Configuration I.}
\label{fig:z3field}
\end{center}
\end{figure}

The waterfall effective mass parameters for $N=3$ are given by
\bea
m^2_{\chi,1} &=& m^2_\chi -\mu^2 \cos\Big(\frac{\phi}{f}+\frac{2\pi}{3}\Big) \nonumber \\
&=&m^2_\chi +\frac{1}{2}\mu^2\bigg(\cos\Big(\frac{\phi}{f}\Big)+\sqrt{3}\sin\Big(\frac{\phi}{f}\Big)\bigg),  \label{z3m1} \\
m^2_{\chi,2} &=& m^2_\chi -\mu^2 \cos\Big(\frac{\phi}{f}+\frac{4\pi}{3}\Big) \nonumber \\
&=& m^2_\chi +\frac{1}{2}\mu^2\bigg(\cos\Big(\frac{\phi}{f}\Big)-\sqrt{3}\sin\Big(\frac{\phi}{f}\Big)\bigg), \label{z3m2} \\
m^2_{\chi,3}  &=&m^2_\chi -\mu^2\cos\Big(\frac{\phi}{f}\Big). \label{z3m3}
\eea
Then, we find that the Configuration I for $N=3$ is realized under the following conditions for $x\equiv \cos\Big(\frac{\phi}{f} \Big)/a_0$ and $y\equiv \sin\Big(\frac{\phi}{f} \Big)/a_0$ with $a_0\equiv \frac{m^2_\chi}{\mu^2}<1$ ,
\bea
x<1,\qquad x+\sqrt{3}y >-2, \qquad x-\sqrt{3} y>-2, \qquad x^2+y^2=\frac{1}{a^2_0}>1,
\eea
which corresponds to the circle inside a triangle in the $x-y$ plane due to the $Z_3$ symmetry as shown in Fig.~\ref{fig:z3field}. Outside the above region, one or two of the waterfall mass parameters become negative. 
But, there is no region where  all  the waterfall fields have $m^2_{\chi_k}<0$, so there is no configuration  where the VEVs of all the waterfall fields are nonzero. Moreover,  the regions with two of $m^2_{\chi_k}$'s are negative are not accessible from the Configuration I  (the inflation phase) by the change of the inflation field value, although they are connected to the Configuration II (the waterfall phase) in the region outside the triangle.

We note that we have $m^2_{\chi,1}, m^2_{\chi,2}, m^2_{\chi,3}>0$ along the circle  inside the green triangle, maintaining the unbroken symmetries in the mirror sectors. Then, for instance, the effective mass parameter, $m^2_{\chi,3}$, scans from positive to negative values as the field $\phi$ moves along during inflation from $x<1$ to $x>1$, inducing a waterfall transition.

In Configuration II (the waterfall phase) where some of the waterfall VEVs are  nonzero, for instance, $\langle \chi_3\rangle\neq 0$, the effective potential at tree level  becomes 
\bea
V_{1,\rm eff}(\phi)&=& -\frac{1}{4\lambda} m^4_{\chi,3} +{\rm const}  \nonumber \\
&=&-\frac{\mu^4}{4\lambda}  \Big( \cos \Big(\frac{\phi}{f} \Big)-a_0\Big)^2 +{\rm constant}. 
\eea
Thus, as in the $Z_2$ case, for $a^2_0<1$, for which $m^2_{\chi,3}$ is scannable from positive to negative, there is no stable minimum  of $V_{1,\rm eff}$ in the above Configuration II as in the $Z_2$ case, but we can stabilize it by the inflaton potential ($V_{\rm inf}$) in the true vacuum.

We find that the effective potential for $\phi$ changes from constant to a nontrivial sine-wave form, as we move from the field space with all the waterfall VEVs being zero to that with some of the waterfall VEVs being nonzero. Therefore, the boundary with $m^2_{\chi,3}=0$ is special, dividing two regions of the field configuration.
Due to the inflaton potential with $N=3$, the minimum of the full potential including the inflaton potential ($V_{\rm inf}+V_{1,{\rm eff}}$) appears where some of the waterfall field masses is positive for $\delta=\pi$.  We have some of the waterfall field masses as being positive in the true vacuum, so we choose $\delta=\pi$.

\subsection{The $Z_4$ model}

\begin{figure}[!t]
\begin{center}
\includegraphics[width=0.45\textwidth]{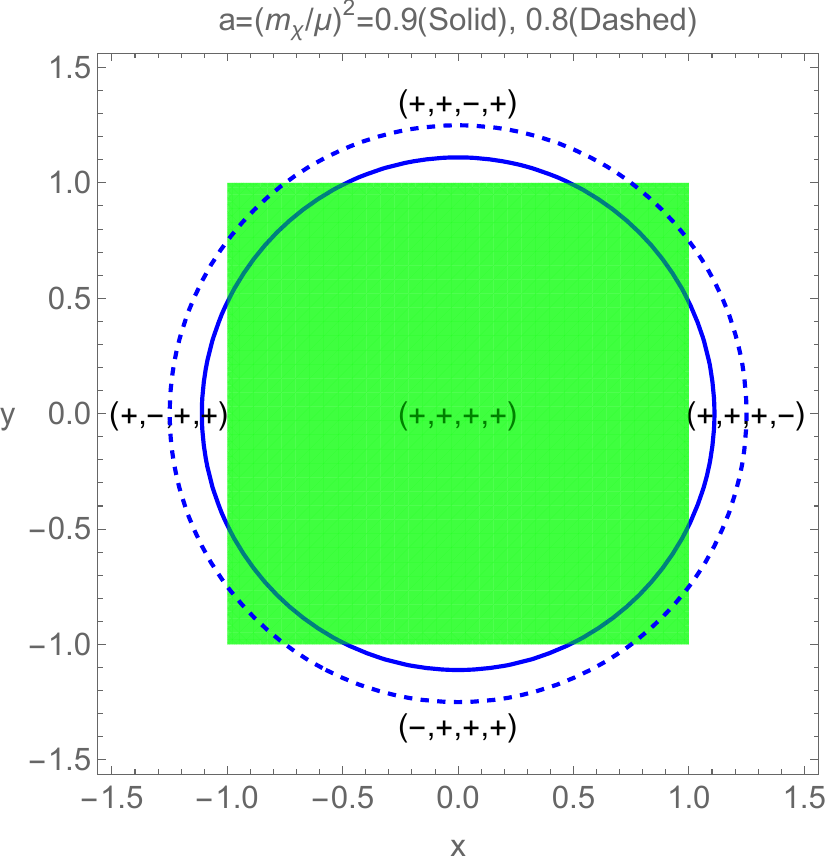}
\caption{Phase space of the field $\phi$ in $x=\cos(\phi/f)/a_0$ versus $y=\sin(\phi/f)/a_0$ for the $Z_4$ case. We chose $a_0\equiv \frac{m^2_\chi}{\mu^2}=0.9, 0.8$ in solid and dashed lines, respectively. Configuration I (the inflation phase) is realized only along the part of solid or dashed lines inside the green square. We have also indicated (${\rm sgn}(m^2_{\chi_1}),{\rm sgn}(m^2_{\chi_2}),{\rm sgn}(m^2_{\chi_3}),{\rm sgn}(m^2_{\chi_4}))$  in Configuration II (the waterfall phase) connected to the Configuration I.}
\label{fig:z4field}
\end{center}
\end{figure}

For the $N=4$ case, we can consider the similar analysis for the effective potential at tree level. 
The waterfall effective mass parameters for $N=4$ are given by
\bea
m^2_{\chi,1} &=& m^2_\chi +\mu^2\sin\Big(\frac{\phi}{f}\Big), \label{z4m1}  \\
m^2_{\chi,2} &=& m^2_\chi +\mu^2\cos\Big(\frac{\phi}{f}\Big), \label{z4m2} \\
m^2_{\chi,3} &=& m^2_\chi -\mu^2\sin\Big(\frac{\phi}{f}\Big), \label{z4m3} \\
m^2_{\chi,4} &=& m^2_\chi -\mu^2\cos\Big(\frac{\phi}{f}\Big). \label{z4m4}
\eea
Thus, the Configuration I with all $\langle \chi_k\rangle= 0$ with $k=1,2,3,4$, is realized under the following conditions for $x\equiv \cos\Big(\frac{\phi}{f} \Big)/a_0$ and $y\equiv \sin\Big(\frac{\phi}{f} \Big)/a_0$,
\bea
|x|<1,\qquad |y|<1, \qquad x^2+y^2=\frac{1}{a^2_0}>1,
\eea
which corresponds to the circle inside a square in the $x-y$ plane due to the $Z_4$ symmetry as shown in Fig. \ref{fig:z4field}. 
Outside the above region, one or two of the waterfall mass parameters become negative.
But, there is no region where more than two waterfall mass parameters are negative, so there is no configuration where the VEVs of all the waterfall VEVs are nonzero. Moreover,  the regions with two of $m^2_{\chi_k}$'s are negative are not accessible from the Configuration I (the inflation phase) by the change of the inflation field value, although they are connected to the Configuration II (the waterfall phase) in the region outside the square.

For $|y|<1$ and $|x|>1$, we have $m^2_{\chi,1}, m^2_{\chi_2},m^2_{\chi,3}, m^2_{\chi,4}>0$ along the circle inside the green square, maintaining the unbroken symmetries in the mirror sectors. Then, for instance, the effective mass parameter, $m^2_{\chi,4}$, scans from positive to negative values, as the field $\phi$ moves along during inflation from $|x|<1$ to $|x|>1$, inducing a waterfall transition.

On the other hand, in Configuration II where only some of the waterfall VEVs  are nonzero, for instance, $\langle \chi_4\rangle\neq 0$, the effective potential at tree level is given by
\bea
V_{1,\rm eff}(\phi)&=&- \frac{\mu^4}{4\lambda} m^4_{\chi,4}+{\rm constant} \nonumber \\
&=&  -\frac{\mu^4}{4\lambda} \bigg(\cos\Big(\frac{\phi}{f}\Big)-a_0\bigg)^2  +{\rm constant}.
\eea
Therefore, as in the $Z_2$ or $Z_3$ cases,  for $a^2_0<1$, for which $m^2_{\chi,4}$ is scannable from positive to negative, there is no stable minimum of $V_{1,\rm eff}$ alone for the above Configuration II, but we can also stabilize it by the inflaton potential $V_{\rm inf}$ in the true vacuum.

\subsection{Effects of mixing masses for waterfall transitions}

We consider the case with nonzero mixing masses for the waterfall fields.

\subsubsection{The $Z_2$ model}

For $N=2$, the mass matrix for the waterfall fields with a mixing mass is given by
\bea
{\cal M}^2_{N=2}=\left( \begin{array}{cc} m^2_{\chi,1} & 2\alpha^2 \vspace{0.2cm} \\  2\alpha^2 & m^2_{\chi,2} \end{array} \right)
\eea
where $m^2_{\chi,1}, m^2_{\chi,2}  $ are given in eqs.~(\ref{z2m1}) and (\ref{z2m2}).
Then, the inflaton-dependent effective mass eigenvalues are given by
\bea
M^2_{1,2}(\phi) &=& \frac{1}{2}\Big( m^2_{\chi,1} +m^2_{\chi,2} \pm \sqrt{(m^2_{\chi,1} -m^2_{\chi,2} )^2+16\alpha^4} \bigg) \nonumber \\
&=& m^2_\chi \pm \sqrt{\mu^4\cos^2\Big(\frac{\phi}{f}\Big) +4\alpha^4}. \label{mixingZ2}
\eea
As a result, only one of the waterfall fields changes the sign of its squared mass during inflation, so there is a transition from Configuration I to Configuration II, as in the case with $\alpha=0$.

\subsubsection{The $Z_3$ model}

For $N=3$, the mass matrix for the waterfall fields with mixing masses is given by
\bea
{\cal M}^2_{N=3}=\left( \begin{array}{ccc} m^2_{\chi,1} & \alpha^2 & \alpha^2  \vspace{0.2cm} \\   \alpha^2 & m^2_{\chi,2} & \alpha^2  \vspace{0.2cm}  \\ \alpha^2 &  \alpha^2 & m^2_{\chi,3} \end{array} \right)
\eea
where $m^2_{\chi,1}, m^2_{\chi,2}, m^2_{\chi,3} $ are given in eqs.~(\ref{z3m1})-(\ref{z3m3}). Then, the inflaton-dependent effective mass eigenvalues are given by
\bea
M^2_1&=&m^2_\chi + 2\bigg(\alpha^4 +\frac{1}{4}\mu^4\bigg)^{1/2} \cos\bigg(\frac{\theta+4\pi}{3}\bigg),  \label{mixingZ3-1}\\
M^2_2&=&m^2_\chi + 2\bigg(\alpha^4 +\frac{1}{4}\mu^4\bigg)^{1/2} \cos\bigg(\frac{\theta}{3}\bigg) ,  \label{mixingZ3-2} \\
M^2_3&=&m^2_\chi + 2\bigg(\alpha^4 +\frac{1}{4}\mu^4\bigg)^{1/2} \cos\bigg(\frac{\theta+2\pi}{3}\bigg),  \label{mixingZ3-3}
\eea
with
\bea
\cos\theta= \frac{\alpha^6-\frac{1}{8}\mu^6 \cos\big(\frac{3\phi}{f}\big)}{\big(\alpha^4+\frac{1}{4}\mu^4\big)^{3/2}}. \label{theta}
\eea

We note that for $\alpha=0$, we take one of the solution of eq.~(\ref{theta}) for $\cos\big(\frac{\theta}{3}\big)$  to be $-\cos\big(\frac{3\phi}{f}\big)$, recovering the waterfall field masses with $\alpha=0$ in Section 3.2.

For $\alpha\neq 0$, there is a boundary of the inflaton field for the waterfall transition along which ${\rm det}({\cal M}^2_{N=3})$ vanishes, so the phase diagrams for the inflation vacua can be obtained in a similar way as in the case with $\alpha=0$.

\section{Hybrid natural inflation with loop corrections}

For natural hybrid inflation, we consider the waterfall transition from Configuration I (without VEVs) to Configuration II (with some of nonzero VEVs). We discuss the cases with $Z_2$, $Z_3$ and $Z_4$, in order.

\subsection{The $Z_2$ model}

For the initial condition satisfying $\cos\big(\frac{\phi}{f}\big)<\frac{m^2_\chi}{\mu^2}$, we have both waterfall fields  stable during inflation, because of $m^2_{\chi,1}, m^2_{\chi,2}>0$ in the $Z_2$ case. Then, inflation ends when $m^2_{\chi,2}$ changes its sign to a negative value, namely, at $\phi=\phi_c$ with $\phi_c=f {\rm arccos}(m^2_\chi/\mu^2)$, so the waterfall transition occurs along the $\chi_2$ direction. 
 
During inflation, we obtain the CW potential for the $Z_2$ case as
\bea
V_{{\rm CW},Z_2}&=& \frac{1}{16\pi^2} m^2_\chi M^2_*- \frac{1}{64\pi^2} \sum_{k=1}^2  m^4_{\chi,k} \ln \frac{e^{\frac{1}{2}}M^2_*}{m^2_{\chi,k}} \nonumber \\
&\simeq& \frac{1}{16\pi^2} m^2_\chi M^2_*- \frac{1}{32\pi^2}\Big(m^4_\chi +\mu^4 \cos^2\bigg(\frac{\phi}{f}\bigg) \Big)\ln \frac{M^2_*}{m^2_\chi}.
\eea
When the mixing mass term is included as in eq.~(\ref{mixingZ2}), we only need to replace $m^4_\chi $ by $m^4_\chi +\alpha^4$ in the second line in the above result. In this case, the CW potential gives rise to additional terms in the inflaton potential with logarithmic divergence, which depend on the waterfall field couplings to the inflaton. Those log terms,  $-\frac{1}{64\pi^2}\sum_{k=1}^2m^4_{\chi,k}\ln (e^{\frac{1}{2}}\,m^2_\chi/m^2_{\chi,k})$, can be ignored as compared to the $\ln(M^2_*/m^2_\chi)$ term, for $M_*\sim 4\pi f\gg m_\chi\sim m_{\chi,k}$  during inflation, as will be discussed below in more detail.

In the $Z_2$ case, the one-loop corrected inflaton potential is given by
\bea
V_{Z_2,{\rm  1L}}\simeq V_{0,R}-{\tilde \Lambda}^4 \cos\bigg( \frac{2\phi}{f}\bigg)
 \label{VeffZ2}
\eea
where $V_{0,R}$ is the renormalized cosmological constant term, given by
\bea
V_{0,R}&=&V_0+\frac{1}{16\pi^2}\, m^2_\chi M^2_* -\frac{1}{32\pi^2}\,\Big(m^4_\chi +\frac{1}{2}\mu^4\Big)\ln \frac{M^2_*}{m^2_\chi}, \\
{\tilde \Lambda}^4&=& \Lambda^4+\frac{\mu^4}{64\pi^2}\ln \frac{M^2_*}{m^2_\chi},
\eea
 and we took $\delta=\pi$ in the tree-level inflaton potential.  As the resultant CW potential takes the approximately same form as for the tree-level potential, but the overall factor in the inflaton potential is shifted by the loop corrections. 
 
The $Z_2$ model is identical to the one in Ref.~\cite{Lee:2022fkd,Lee:2023dcy}, up to the definition of the inflaton by $\phi/f\to 2\phi/f+\pi$.
Consequently, as far as the mass parameters in the waterfall field sector satisfies $H^2_I\lesssim \mu^2 \sim m^2_{\chi}\lesssim 4\sqrt{2}\pi \Lambda^2$, the waterfall fields remain decoupled and affect the inflaton mass little. For the pseudo-Goldstone inflaton coming from a QCD-like theory with $\Lambda\lesssim f$,
we can maintain a natural hierarchy of scales in our model, 
\bea
|m^2_\phi|=\frac{4\Lambda^4}{f^2}\ll H^2_I\ll \mu^2\sim m^2_\chi \ll 4\sqrt{2}\pi \Lambda^2 \ll f^2 .
\eea
Then, when the waterfall fields have masses of order $m^2_\chi\gg H^2_I$ during inflation, we can safely decouple the waterfall fields without spoiling the flat direction along the $\phi$ field,  so focus on the dynamics of the $\phi$ field during inflation. 
We note that the duration time of the waterfall transition \cite{Lee:2023dcy} is given by $t_*\simeq \ln(32\pi^2/\lambda_\chi)/(2m_\chi)$, which is given by $t_*\simeq 14/m_\chi-26/m_\chi$ for $\lambda=10^{-20}-10^{-10}$, so  the duration time of the water fall transition or the last stage of inflation is less than the Hubble time for $m_\chi\gg H_I$.

We note that one of the waterfall fields, e.g. $\chi_k$,  turns light near the end of inflation or $\phi=\phi_c$, becoming tachyonic beyond $\phi_c$. The part of the CW potential coming from such a waterfall field $\chi_k$ becomes suppressed by the overall factor $m^4_{\chi,k}$, but we find that the log contributions from $\chi_k$ to the $\eta$ parameter would become IR divergent as some $m^2_{\chi,k}\to 0$. Then, by including the inflaton contributions to the CW potential, we find that the effective mass for the waterfall field is shifted from $m^2_{\chi,k}$ to $m^2_{\chi,k}+\Delta_2$ at loops,  so we can make the resummation of the IR divergence coming from the waterfall field \cite{Espinosa:2017aew}.  Soon after the waterfall field becomes tachyonic, namely, $m^2_{\chi,k}<0$, we assume that the classical rolling of the waterfall field would dominate, inducing a rapid waterfall transition toward the vacuum \cite{Garcia-Bellido:2001dqy,Lee:2023dcy}. 

In the $Z_2$ case,  from eq.~(\ref{fullpot}), the effective mass for the inflaton at tree level is given by
\bea
m^2_{\phi, {\rm eff}}= \frac{4\Lambda^4}{f^2}\,\cos\Big(\frac{2\phi}{f}\Big) +(\chi^2_2-\chi^2_1)\,\frac{\mu^2}{2f^2}\,\cos\Big(\frac{\phi}{f}\Big). 
\eea
Then, the waterfall field $\chi_2$ becomes massless at  $\phi=\phi_c$, causing the IR divergence in the CW potential. But,  including the inflaton contribution to the CW potential and making the expansion of the CW potential with respect to $\chi_2$, we find the renormalized effective mass for $\chi_2$ shifted to
\bea
M^2_{\chi,2}=m^2_\chi-\mu^2 \cos\Big(\frac{\phi}{f}\Big) +\Delta_2. \label{shifted}
\eea
with
\bea
\Delta_2= -\frac{\Lambda^4\mu^2}{16\pi^2f^4}\cos\Big(\frac{\phi}{f}\Big)\cos\Big(\frac{2\phi}{f}\Big) \ln\bigg[\frac{f^2 m^2_\chi}{4\Lambda^4 \big|\cos\big(\frac{2\phi}{f}\big)\big|}\bigg]. \label{resum}
\eea
Therefore, we can resum the contributions coming from the waterfall field $\chi_2$ consistently in the loop expansion, so there is no IR divergence in the improved CW potential. But, the detailed analysis of the resummed CW potential at higher loops is beyond the scope of our work.

We assume that the inflaton starts rolling near $2\phi/f \simeq -\pi$ (at the top of the local maximum of the potential) and increases during inflation until it reaches $\phi_c$. 
For the effective renormalized inflaton potential for the $Z_2$ case, including the log terms explicitly at one-loop order, we obtain the slow-roll parameters for $\mu^4\lesssim {\tilde\Lambda}^4\ll V_{0,R}$, as follows,
\bea
\epsilon&\simeq& \frac{8M^2_P{\tilde\Lambda}^8}{f^2 V_{0,R}^2} \sin^2\Big(\frac{\phi}{f}\Big)\bigg[ \cos\Big(\frac{\phi}{f}\Big)\bigg(1+\frac{\mu^4}{128\pi^2{\tilde\Lambda}^4} \ln \bigg(\frac{m^4_\chi}{m^4_\chi-\mu^4\cos^2\big(\frac{\phi}{f}\big)}\bigg) \bigg) \nonumber \\
&&+\frac{\mu^4}{128\pi^2 {\tilde\Lambda}^4} \bigg(1+\frac{m^2_\chi}{\mu^2}\cos\Big(\frac{\phi}{f}\Big) \ln \bigg(\frac{m^2_\chi-\mu^2\cos\big(\frac{\phi}{f}\big)}{m^2_\chi+\mu^2 \cos\big(\frac{\phi}{f}\big)}\bigg) \bigg) \bigg]^2,  \label{epz20} \\
\eta &\simeq & \frac{4M^2_P{\tilde\Lambda}^4}{f^2 V_{0,R}} \,\bigg[\cos\Big(\frac{2\phi}{f}\Big)\bigg(1+\frac{\mu^4}{128\pi^2{\tilde\Lambda}^4} \ln \bigg(\frac{m^4_\chi}{m^4_\chi-\mu^4 \cos^2\big(\frac{\phi}{f}\big)} \bigg) \bigg)\nonumber \\
&&+ \frac{\mu^4}{128\pi^2{\tilde\Lambda}^4}  \bigg(1+\frac{m^2_\chi}{\mu^2} \cos\Big(\frac{\phi}{f}\Big) \ln \bigg(\frac{m^2_\chi-\mu^2\cos\big(\frac{\phi}{f}\big)}{m^2_\chi+\mu^2\cos\big(\frac{\phi}{f}\big)} \bigg) \bigg) \bigg]. \label{etaz20}
\eea
Here, we note that the $\epsilon$ parameter is finite near the waterfall transition, namely, $\phi=\phi_c$, whereas the $\eta$ parameter becomes IR divergent at  $\phi=\phi_c$.
However, as far as $m^2_\chi\gtrsim \mu^2\cos\big(\frac{\phi}{f}\big)$ during inflation and assuming $\mu^4\lesssim {\tilde\Lambda}^4$,  the loop corrections can be ignored due to the small coefficients of the logarithms. Moreover, towards the waterfall transition along $\chi_2$ when inflation ends, we need to replace $m^2_{\chi,2}=m^4_\chi-\mu^4 \cos^2\big(\frac{\phi}{f}\big)$ by $M^2_{\chi,2}=m^2_{\chi,2}+3\lambda \chi^2_2+\Delta_2$, so there is no IR divergence at $\phi=\phi_c$ and $\chi_1=\chi_2=0$, in the limit of $m^2_{\chi,2}\to 0$, and  the log terms can be still ignored.  

Setting $\phi=\phi_c$, we compute the loop-corrected slow-roll parameters in eqs.~(\ref{epz20}) and (\ref{etaz20}) at the waterfall transition with the shifted mass for the waterfall field $\chi_2$ in eq.~(\ref{shifted}), as follows,
\bea
\epsilon &\simeq& \frac{8M^2_P{\tilde\Lambda}^8}{ f^2 V^2_{0,R}} \sin^2\Big(\frac{\phi_c}{f}\Big)\cos^2\Big(\frac{\phi_c}{f}\Big)(1+\delta_1)^2, \\
\eta &=& \frac{4M^2_P{\tilde\Lambda}^4}{f^2 V_{0,R}} \,\cos\Big(\frac{2\phi_c}{f}\Big) (1 +\delta_2).
\eea
Here,
\bea
\delta_1 &=&-\frac{\ln(2)\mu^4}{64\pi^2\Lambda^4}, \\
\delta_2 &=&\frac{\mu^4}{128\pi^2 {\tilde\Lambda}^4\cos\big(\frac{2\phi_c}{f}\big)}\bigg[(2+\ln(2)-\frac{m^4_\chi}{\mu^4}(2+\ln(8))\nonumber \\
&& +\bigg(\frac{m^4_\chi}{\mu^2}-1\bigg) \ln\Big(\frac{16\pi^2 f^4}{{\tilde\Lambda}^4K} \Big)\bigg],
\eea
with 
\bea
K=\bigg|\cos\Big(\frac{2\phi_c}{f}\Big)\bigg|\ln\bigg[\frac{f^2m^2_\chi}{4{\tilde\Lambda}^4\big|\cos\big(\frac{2\phi_c}{f}\big)\big|} \bigg].
\eea
Then, for $\frac{m^2_\chi}{\mu^2}\neq \frac{1}{2}$, $\cos\big(\frac{2\phi_c}{f}\big)=\frac{2m^2_\chi}{\mu^2}-1$ is nonzero, so the logarithm in the loop correction $\delta_2$ is finite. Therefore, as far as $\mu^4, m^4_\chi\lesssim {\tilde\Lambda}^4$, we can safely ignore the loop corrections to the slow-roll parameters for the inflaton near the waterfall transition.
As soon as $m^2_{\chi,2}<0$, a nonzero field value for the waterfall field $\chi_2$ is developed, so the waterfall transition occurs rapidly, ending the inflation.

As a result, during inflation, we can approximate the slow parameters further to the one at the tree level with $V_0$ and $\Lambda^4$ being replaced by the renormalized parameters, $V_{0,R}$ and ${\tilde\Lambda}^4$, respectively, as follows,
\bea
\epsilon&\simeq& \frac{2M^2_P{\tilde\Lambda}^8}{f^2 V_{0,R}^2} \sin^2\Big(\frac{2\phi}{f}\Big),  \label{epz2}  \\
\eta &\simeq & \frac{4M^2_P{\tilde\Lambda}^4}{f^2 V_{0,R}} \,\cos\Big(\frac{2\phi}{f}\Big),  \label{etaz2}
\eea
and he number of efoldings is also obtained by
\bea
N_e&=&\frac{1}{M_P}\int^{\phi_c}_{\phi_*} \frac{{\rm sgn(V')}}{\sqrt{2\epsilon}}\,d\phi \nonumber \\
&\simeq &\frac{f^2 V_{0,R}}{4 M^2_P{\tilde\Lambda}^4}\bigg[\ln \cot\Big(\frac{\phi_*}{f}\Big)\bigg]  
-(\phi_*\to \phi_c).
\eea
Here, $\phi_*, \phi_c$ are the inflaton field values at the horizon exit and at the end of inflation, respectively. 
We can further approximate the number of efoldings as
\bea
N_e\simeq \frac{f^2 V_{0,R}}{4M^2_P {\tilde\Lambda}^4}  \ln \left(\frac{\cot\Big(\frac{\phi_c}{f}\Big)}{\cot\Big(\frac{\phi_*}{f}\Big)}\right).
\label{efoldsz2}
\eea
Here, we note that $-\log\big(\cot\big(\frac{\phi}{f}\big)\big)=\frac{1}{2}\ln\big[\big(1-\cos\big(\frac{2\phi}{f}\big)\big)/\big(1+\cos\big(\frac{2\phi}{f}\big))\big]>0$ for $\cos\big(\frac{2\phi}{f}\big)<0$ during the evolution of the inflaton.

As a result, the spectral index and the tensor-to-scalar ratio are given by
\bea
n_s &=& 1+2\eta_* -6\epsilon_*,  \label{tiltz2} \\
r&=& 16\epsilon_*
\eea
Moreover, the CMB normalization, $A_s\simeq \frac{1}{24\pi^2} \frac{V_{0,R}}{\epsilon_* M^4_P}\simeq 2.1.\times 10^{-9}$, gives rise to
\bea
r=3.2\times 10^7\cdot \frac{V_{0,R}}{M^4_P}.
\eea
To fit the Planck data, we choose $\epsilon_*\ll |\eta_*|$. Then, the critical value of the inflaton satisfies $|2\phi_*/f+\pi |\lesssim |2\phi_c/f +\pi|\lesssim 1$ for a sufficient number of efoldings $N=40-60$ to solve the horizon problem. 

The observed spectral index is given by $n_s=0.9649\pm 0.0044 $ from Planck \cite{Planck}, $n_s=0.9709\pm0.0038$ from Planck$+$ACT \cite{ACT}, and $n_s=0.9743\pm0.0034$  from Planck$+$ACT$+$LB \cite{ACT}. On the other hand, the bound on the tensor-to-scalar ratio from the combined Planck and Keck data is given by $r<0.036$ at $95\%$ CL \cite{keck}, setting the upper bound on the Hubble scale during inflation  as $H_I < 4.6\times 10^{13}\,{\rm GeV}$.

It is convenient to rewrite the spectral index in eq.~(\ref{tiltz2}), the number of efoldings in eq.~(\ref{efoldsz2})  and the Hubble scale, as follows,
\bea
n_s &\simeq & 1+2\eta_* -\frac{3 f^2 \eta^2_*}{4M^2_P} \, \tan^2\Big(\frac{2\phi_*}{f}\Big), \\
N_e&=& \frac{1}{\eta_*} \bigg[\cos\Big(\frac{2\phi_*}{f}\Big) \bigg]\, \ln \left(\frac{\cot\Big(\frac{\phi_c}{f}\Big)}{\cot\Big(\frac{\phi_*}{f}\Big)} \right), \\
\frac{H_I}{f} &=&1.4\times 10^{-4}\, \left|\eta_* \tan\Big(\frac{2\phi_*}{f}\Big) \right|. \label{Hubblez2}
\eea 
Here, we note that $\eta_*$ is related to the physical mass scales by $\eta_*=|m^2_\phi|/(3H^2_I) \cos\Big(\frac{2\phi_*}{f}\Big)$ with the squared inflaton mass being $|m^2_\phi|=\frac{4{\tilde\Lambda}^4}{f^2}$. From $|\eta_*|\ll 1$, we need $|m^2_\phi|\ll 3H^2_I$ or ${\tilde\Lambda}^4/V_{0,R}\ll \frac{f^2}{4M^2_P}$, which is consistent with ${\tilde\Lambda}^4\ll V_{0,R}$ for $f<2M_P$. We also find that the $\epsilon_*$ contribution to the spectral index $n_s$ is negligible for $f<2M_P$ because it is suppressed by $\eta^2_*$. For instance, for $\cos\big(\frac{2\phi_*}{f}\big)=-0.90(-0.80)$ and $n_s=0.9649$, we get the Hubble scale during inflation as $H_I/f\simeq 1.2(1.8)\times 10^{-6}$. Then,  for $\cos\big(\frac{2\phi_*}{f}\big)=-0.90(-0.80)$, taking $m_\chi=100H_I$ and $\mu/m_\chi=1.5(1.3)$, we find that the numerical values of the loop corrections to the slow-roll parameters at the waterfall transition in eqs.~(\ref{epz20}) and (\ref{etaz20}) are $\delta_1=-8\times 10^{-5} (-10^{-4})$ and $\delta_2=2.5\times 10^{-3}(4.3\times 10^{-3})$, so it is sufficient to take the tree-level slow-roll parameters in eqs.~(\ref{epz2}) and (\ref{etaz2}).

\begin{figure}[!t]
\begin{center}
\includegraphics[width=0.45\textwidth]{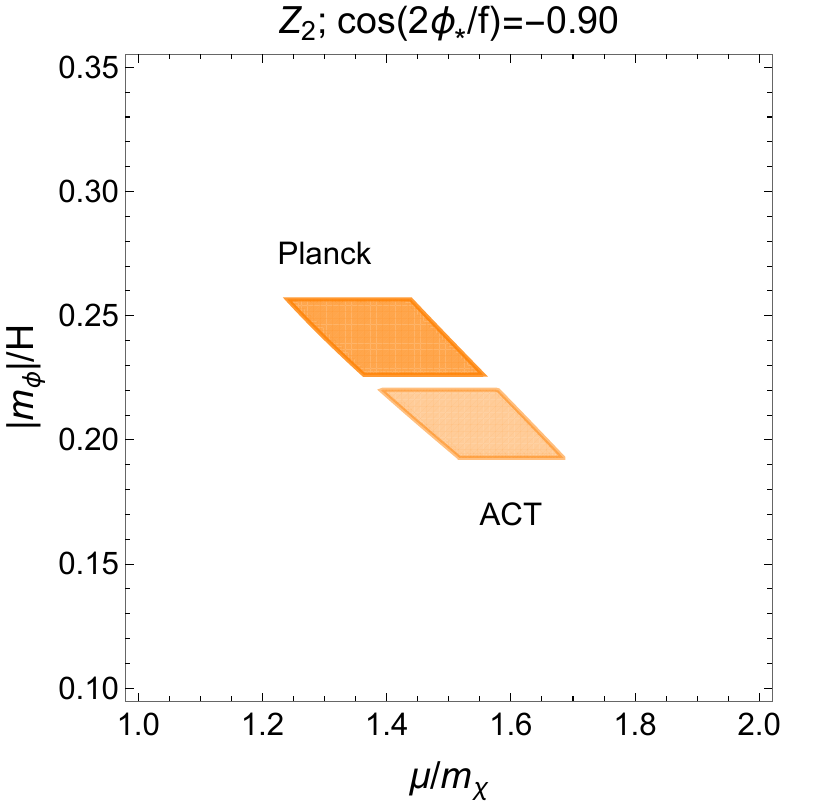} \,\, \includegraphics[width=0.45\textwidth]{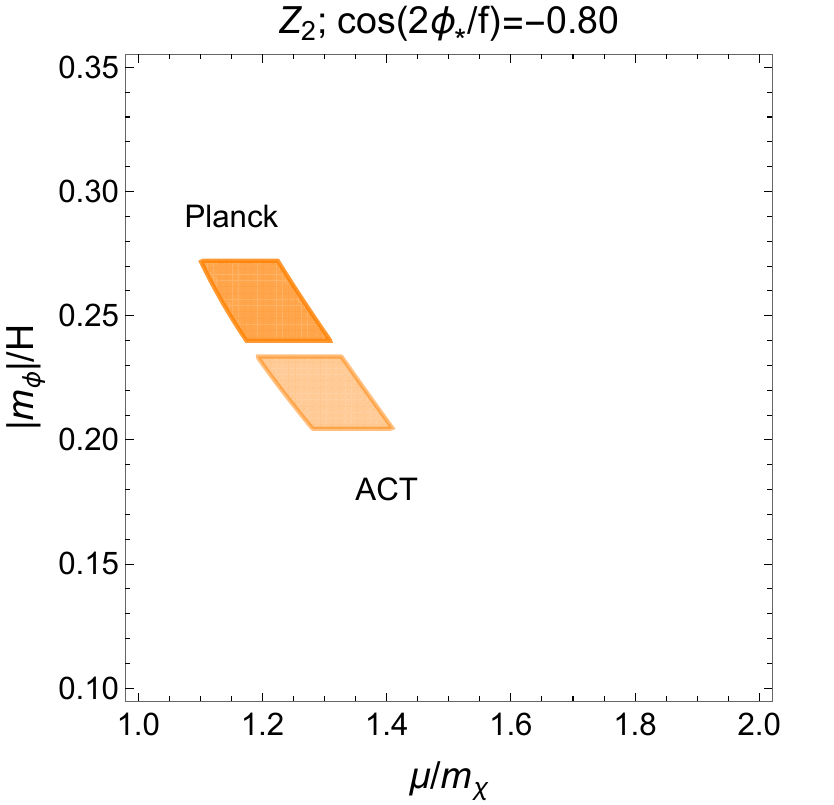}  \vspace{0.5cm} \\ 
\includegraphics[width=0.45\textwidth]{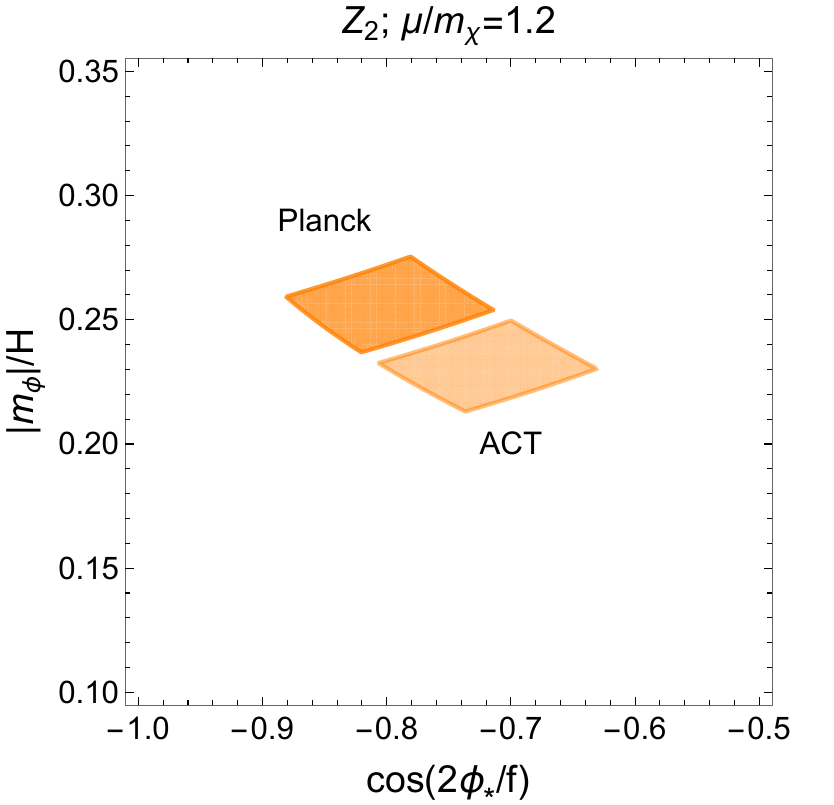} \,\, \includegraphics[width=0.45\textwidth]{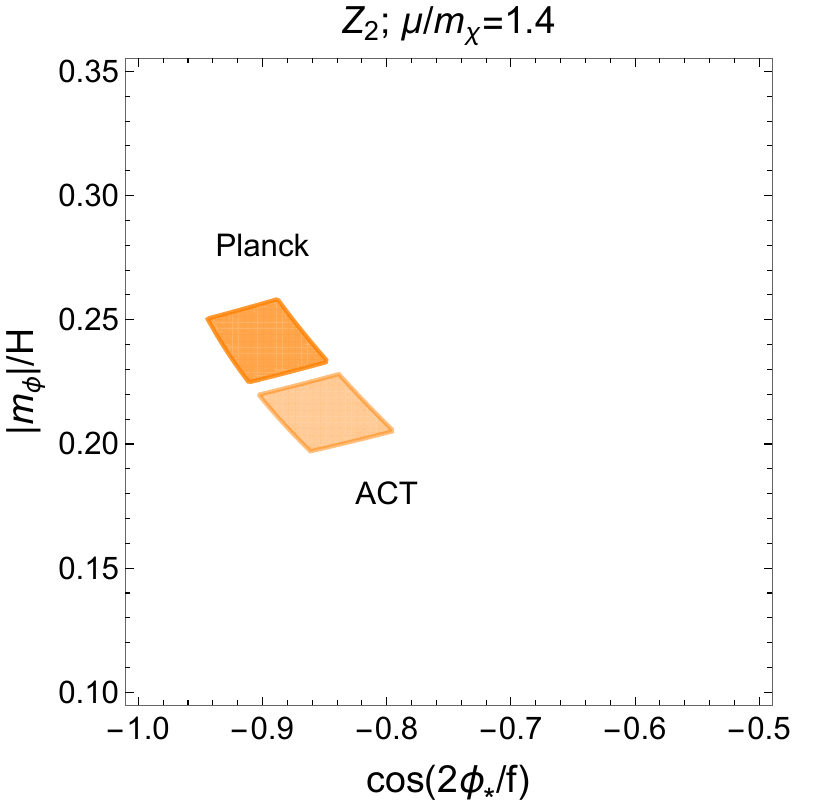} 
\caption{Parameter space for the $Z_2$ model at the one-loop order in orange, being consistent with Planck data and $N=40-60$. Dark colors correspond to the Planck $1\sigma$ band and light colors correspond to the ACT $1\sigma$ band.}
\label{fig:z2}
\end{center}
\end{figure}

In Fig.~\ref{fig:z2}, we depict the parameter space for a successful inflation with $N=40-60$ at the one-loop order in orange, being consistent with Planck $1\sigma$ \cite{Planck} in thick colors and ACT $1\sigma$ \cite{ACT} in light colors. In the upper panel, we show the parameter space for $a_0=\mu/m_\chi$ vs $|m^2_\phi|/H_I$, with a varying the inflation field value at the horizon exit, $\cos(2\phi_*/f)=-0.90, -0.80$ on left and right, respectively. On the other hand, in the lower panel, we also show the parameter space for $\cos(2\phi_*/f)$ vs $|m^2_\phi|/H_I$, with $\mu/m_\chi=1.2, 1.4$ on left and right, respectively.

\subsection{The $Z_3$ and $Z_4$ models}

For the initial condition satisfying $\cos\big(\frac{\phi}{f}\big)<\frac{m^2_\chi}{\mu^2}$, $\cos\big(\frac{\phi}{f}\big)+\sqrt{3}\sin\big(\frac{\phi}{f}\big)>-\frac{2m^2_\chi}{\mu^2} $ and $\cos\big(\frac{\phi}{f}\big)-\sqrt{3}\sin\big(\frac{\phi}{f}\big)>-\frac{2m^2_\chi}{\mu^2} $, we have all the waterfall fields stable during inflation due to $m^2_{\chi,i}>0$ with $i=1,2,3$ in the $Z_3$ case.
Then, as the inflaton moves from $\cos\big(\frac{\phi}{f}\big)<\frac{m^2_\chi}{\mu^2}$ to $\cos\big(\frac{\phi}{f}\big)>\frac{m^2_\chi}{\mu^2}$,  $m^2_{\chi,3}$ changes its sign to a negative value, namely, at $\phi=\phi_c$ with $\phi_c=f {\rm arccos}(m^2_\chi/\mu^2)$, so the waterfall transition takes place along $\chi_3$ and inflation ends. 

Similarly, if the inflaton moves from $\cos\big(\frac{\phi}{f}\big)+\sqrt{3}\sin\big(\frac{\phi}{f}\big)>-\frac{2m^2_\chi}{\mu^2} $ to $\cos\big(\frac{\phi}{f}\big)+\sqrt{3}\sin\big(\frac{\phi}{f}\big)>-\frac{2m^2_\chi}{\mu^2} $, $m^2_{\chi,1}$ changes its sign to a negative value, namely, at $\phi=\phi_c-\frac{2\pi}{3}\, f$ with $\phi_c=f {\rm arccos}(m^2_\chi/\mu^2)$, so the waterfall transition takes place along $\chi_1$. If the inflaton moves from $\cos\big(\frac{\phi}{f}\big)-\sqrt{3}\sin\big(\frac{\phi}{f}\big)>-\frac{2m^2_\chi}{\mu^2} $ to $\cos\big(\frac{\phi}{f}\big)-\sqrt{3}\sin\big(\frac{\phi}{f}\big)<-\frac{2m^2_\chi}{\mu^2} $, $m^2_{\chi,2}$ changes its sign to a negative value, namely, at $\phi=\phi_c-\frac{4\pi }{3}\, f$ with $\phi_c=f {\rm arccos}(m^2_\chi/\mu^2)$, so the waterfall transition takes place along $\chi_2$. 

During inflation, we obtain the CW potential for the $Z_3$ case as
\bea
V_{{\rm CW},Z_3}&=& \frac{3}{32\pi^2} m^2_\chi M^2_*- \frac{1}{64\pi^2} \sum_{k=1}^3  m^4_{\chi,k} \ln \frac{e^{\frac{1}{2}}M^2_*}{m^2_{\chi,k}} \nonumber \\
&\simeq &\frac{3}{32\pi^2} m^2_\chi M^2_*- \frac{1}{64\pi^2} \Big(3m^4_\chi +\frac{3}{2}\mu^4\Big)\ln \frac{M^2_*}{m^2_\chi}.
\eea
When the mixing mass terms are included as in eqs.~(\ref{mixingZ3-1})-(\ref{mixingZ3-3}), we only need to replace $\mu^4$ by $\mu^4 +4\alpha^4$ in the second line in the above result. 
For the $Z_3$ case, the CW potential is insensitive to the waterfall field couplings, due to the mild logarithmic dependence of the loop corrections on the inflaton, unlike the $Z_2$ case. 
We note that some of the log terms,  $-\frac{1}{64\pi^2}\sum_{k=1}^3m^4_{\chi,k}\ln (e^{\frac{1}{2}}\,m^2_\chi/m^2_{\chi,k})$, contribute to the inflaton potential, but those log terms are sub-dominant as $m_{\chi,k}\sim m_\chi$  during inflation.

As in the $Z_2$ case, In the $Z_3$ case,  from eq.~(\ref{fullpot}), we obtain the effective mass for the inflaton at tree level as
\bea
m^2_{\phi, {\rm eff}}&=& \frac{9\Lambda^4}{f^2}\,\cos\Big(\frac{3\phi}{f}\Big) +\sqrt{3}(\chi^2_2-\chi^2_1)\,\frac{\mu^2}{4f^2}\,\sin\Big(\frac{\phi}{f}\Big) \nonumber \\
&&+(2\chi^2_3-\chi^2_1-\chi^2_2)\frac{\mu^2}{4f^2} \cos\Big(\frac{\phi}{f}\Big). 
\eea
Then, the waterfall field $\chi_3$ becomes massless at  $\phi=\phi_c$, causing the IR divergence in the CW potential. But,  we can include the inflaton contribution to the CW potential and make the expansion of the CW potential with respect to $\chi_3$. Then, we find the renormalized effective mass for $\chi_3$ shifted to
\bea
M^2_{\chi,2}=m^2_\chi-\mu^2 \cos\Big(\frac{\phi}{f}\Big) +\Delta_3.
\eea
with
\bea
\Delta_3= -\frac{9\Lambda^4\mu^2}{64\pi^2f^4}\cos\Big(\frac{\phi}{f}\Big)\cos\Big(\frac{3\phi}{f}\Big) \ln\bigg[\frac{f^2 m^2_\chi}{9\Lambda^4 \big|\cos\big(\frac{3\phi}{f}\big)\big|}\bigg]. \label{resum}
\eea
Therefore, we can make a resummation of the loop contributions coming from the waterfall field $\chi_3$ consistently, so there is no IR divergence in the improved CW potential as in the $Z_2$ case. 
We also note that for $\mu^4\lesssim {\Lambda}^4$,  the loop corrections during inflation can be ignored due to the  small coefficients of the logarithms. Moreover,  when inflation ends due to the waterfall transition along $\chi_3$, we need to replace $m^2_{\chi,3}$ by $M^2_{\chi,3}=m^2_{\chi,3}+3\lambda \chi^2_3+\Delta_3$, so there is no IR divergence at $\phi=\phi_c$ and $\chi_1=\chi_2=\chi_3=0$, in the limit of $m^2_{\chi,3}\to 0$, and the log terms can be still ignored until the waterfall transition as in the $Z_2$ case. 

In the $Z_3$ case, the one-loop corrected inflaton potential is given by
\bea
V_{Z_3,{\rm 1L}}= V_{0,R}-\Lambda^4 \cos\bigg( \frac{3\phi}{f}\bigg),
\eea
with
\bea
V_{0,R}=V_0 +\frac{3}{32\pi^2} m^2_\chi M^2_*- \frac{1}{64\pi^2} \Big(3m^4_\chi +\frac{3}{2}\mu^4\Big)\ln \frac{M^2_*}{m^2_\chi}.
\eea
In this case, the coefficient of the resultant inflaton potential does not get renormalized, unlike the $Z_2$ case, so it is insensitive to the couplings to the waterfall fields. Then, we only need a milder hierarchy between the Hubble scale and the QCD-like scale $\Lambda$/the inflaton decay constant $f$, 
\bea
|m^2_\phi|=\frac{9\Lambda^4}{f^2}\ll H^2_I \ll \mu^2\sim m^2_\chi \ll \Lambda^2 \ll f^2 .
\eea
As in the $Z_2$ case, the waterfall fields can have masses of order $m^2_\chi\gg H^2_I$ during inflation leaving the $\phi$ field direction flat, so we can safely decouple the waterfall fields and focus on the dynamics of the $\phi$ field during inflation.

For the $Z_4$ model, for the initial condition satisfying $-\frac{m^2_\chi}{\mu^2}<\cos\big(\frac{\phi}{f}\big)<\frac{m^2_\chi}{\mu^2}$ and $-\frac{m^2_\chi}{\mu^2}<\sin\big(\frac{\phi}{f}\big)<\frac{m^2_\chi}{\mu^2}$, we have all the waterfall fields stable during inflation due to $m^2_{\chi,i}>0$ with $i=1,2,3,4$ in the $Z_4$ case. 
Then, as the inflaton goes outside the region for the stable waterfall fields, the waterfall transition occurs and the inflation ends.
For instance, the inflation can move from $\cos\big(\frac{\phi}{f}\big)<\frac{m^2_\chi}{\mu^2}$ to $\cos\big(\frac{\phi}{f}\big)>\frac{m^2_\chi}{\mu^2}$ while $-\frac{m^2_\chi}{\mu^2}<\sin\big(\frac{\phi}{f}\big)<\frac{m^2_\chi}{\mu^2}$ is satisfied. In this case, $m^2_{\chi,4}$ changes its sign to a negative value, namely, at $\phi=\phi_c$ with $\phi_c=f {\rm arccos}(m^2_\chi/\mu^2)$, so the waterfall transition takes place. 

On the other hand, when the inflation moves $\sin\big(\frac{\phi}{f}\big)<\frac{m^2_\chi}{\mu^2}$ to $\sin\big(\frac{\phi}{f}\big)>\frac{m^2_\chi}{\mu^2}$ while $-\frac{m^2_\chi}{\mu^2}<\cos\big(\frac{\phi}{f}\big)<\frac{m^2_\chi}{\mu^2}$ is satisfied, then $m^2_{\chi,3}$ changes its sign to a negative value, namely, at $\phi=\phi_c$ with $\phi_c=f {\rm arcsin}(m^2_\chi/\mu^2)$, so the waterfall transition takes place similarly.

During inflation, we obtain the CW potential for the $Z_4$ case as
\bea
V_{{\rm CW},Z_4}&=& \frac{1}{8\pi^2} m^2_\chi M^2_*- \frac{1}{64\pi^2} \sum_{k=1}^4  m^4_{\chi,k} \ln \frac{e^{\frac{1}{2}}M^2_*}{m^2_{\chi,k}} \nonumber \\
&\simeq &\frac{1}{8\pi^2} m^2_\chi M^2_*- \frac{1}{64\pi^2} \Big(4m^4_\chi +2\mu^4\Big)\ln \frac{M^2_*}{m^2_\chi}.
\eea
As in the $Z_3$ case, the individual contributions from the waterfall fields to the CW potential are cancelled out, leaving only a mild dependence on the inflaton inside the logarithms. In the $Z_4$ case, the one-loop corrected inflaton potential is given by
\bea
V_{Z_4,{\rm 1L}}= V_{0,R}-\Lambda^4 \cos\bigg( \frac{4\phi}{f}\bigg),
\eea
with
\bea
V_{0,R}=V_0+\frac{1}{8\pi^2} m^2_\chi M^2_*- \frac{1}{64\pi^2} \Big(4m^4_\chi +2\mu^4\Big)\ln \frac{M^2_*}{m^2_\chi}.
\eea
Similarly to the $Z_3$ case, the coefficient of the inflaton potential does not get renormalized, so the resultant inflaton potential is insensitive to the couplings to the waterfall fields.  So, we only need a milder hierarchy between the Hubble scale and the QCD-like scale $\Lambda$/the inflaton decay constant $f$, 
\bea
|m^2_\phi|=\frac{16\Lambda^4}{f^2}\ll H^2_I \ll \Lambda^2 \ll f^2 .
\eea
We also note that some of the log terms,  $-\frac{1}{64\pi^2}\sum_{k=1}^4 m^4_{\chi,k}\ln (e^{\frac{1}{2}}\,m^2_\chi/m^2_{\chi,k})$, contribute to the inflaton potential, but those log terms are sub-dominant as $m_{\chi,k}\sim m_\chi$  during inflation as well as the waterfall transition for the regularized mass for the waterfall field.

In general, for the $Z_N$ cases with $N>2$, the log terms,  $-\frac{1}{64\pi^2}\sum_{k=1}^N m^4_{\chi,k}\ln (e^{\frac{1}{2}}\,m^2_\chi/m^2_{\chi,k})$, also contribute to the inflaton potential, but those log terms after resummation are sub-dominant during inflation and near the waterfall transition.
So, we can maintain the tree-level form of the  inflaton potential approximately, as follows,
\bea
V_{Z_N,{\rm 1L}}= V_{0,R}-\Lambda^4 \cos\bigg( \frac{N\phi}{f}\bigg).
\eea
Here, we note that only $V_{0,R}$ is the renormalized parameter, except for $N=2$.
Then, the corresponding slow-roll parameters for $\mu^4\lesssim \Lambda^4\ll V_{0,R}$, are given by
\bea
\epsilon&\simeq& \frac{N^2 M^2_P\Lambda^8}{2 f^2 V_{0,R}^2}  \sin^2\Big(\frac{N\phi}{f}\Big),  \label{epzn} \\
\eta &\simeq & \frac{N^2 M^2_P\Lambda^4}{f^2 V_{0,R}} \, \cos\Big(\frac{N\phi}{f}\Big), \label{etazn}
\eea
and the number of efoldings is given by
\bea
N_e&\simeq &\frac{f^2 V_{0,R}}{N^2M^2_P \Lambda^4}\bigg[{\rm arctanh}\Big(\cos\Big(\frac{N\phi_c}{f}\Big)\Big) \bigg] -(\phi_c\to \phi_*).
\eea
Here, we note that $-{\rm arctanh}\big(\cos\big(\frac{N\phi}{f}\big)\big)=\frac{1}{2}\ln\big[\big(1-\cos\big(\frac{N\phi}{f}\big)\big)/\big(1+\cos\big(\frac{N\phi}{f}\big))\big]>0$ for $\cos\big(\frac{N\phi}{f}\big)<0$ during the evolution of the inflaton and the result for the number of efolding with $N=2$ is consistent with the tree level result in the $Z_2$ case in eq.~(\ref{efoldsz2}).

Similarly as in the $Z_2$ case, it is convenient to write the spectral index, the number of efoldings  and the Hubble scale, as follows,
\bea
n_s &\simeq & 1+2\eta_* \simeq 1+\frac{2|m^2_\phi|}{3H^2_I}\,\cos\Big(\frac{N\phi_*}{f}\Big), \\
N_e&=& \frac{1}{\eta_*} \, \cos\Big(\frac{N\phi_*}{f}\Big)\bigg[ {\rm arctanh}\Big(\cos\Big(\frac{N\phi_c}{f}\Big)\Big)-{\rm arctanh}\Big(\cos\Big(\frac{N\phi_*}{f}\Big)\Big)\bigg], \\
\frac{H_I}{f} &=& 2.9\times 10^{-4} \,\bigg|\frac{\eta_*}{N} \tan\Big(\frac{N\phi_*}{f}\Big) \bigg| \label{Hubblezn}
\eea 
where the squared inflaton mass is given by $|m^2_\phi|=\frac{N^2\Lambda^4}{f^2}$. From $|\eta_*|\ll 1$, we need $|m^2_\phi|\ll 3H^2_I$ or $\Lambda^4/V_{0,R}\ll \frac{f^2}{N^2M^2_P}$, which is consistent with $\Lambda^4\ll V_{0,R}$ for $f<N M_P$ as in the $Z_2$ case. Then, we note that the $\epsilon_*$ contribution to the spectral index $n_s$ is suppressed by  $\eta^2_*$, so it was neglected for $f<N M_P$ as in the $Z_2$ case.
For instance, taking $\cos\big(\frac{2\phi_*}{f}\big)=-0.90(-0.80)$ and $n_s=0.9649$, we get the Hubble scale during inflation as $H_I/f\simeq 0.82(1.3)\times 10^{-6}$ for $N=3$ and $H_I/f\simeq 6.2(9.5)\times 10^{-7}$ for $N=4$.

\begin{figure}[!t]
\begin{center}
\includegraphics[width=0.45\textwidth]{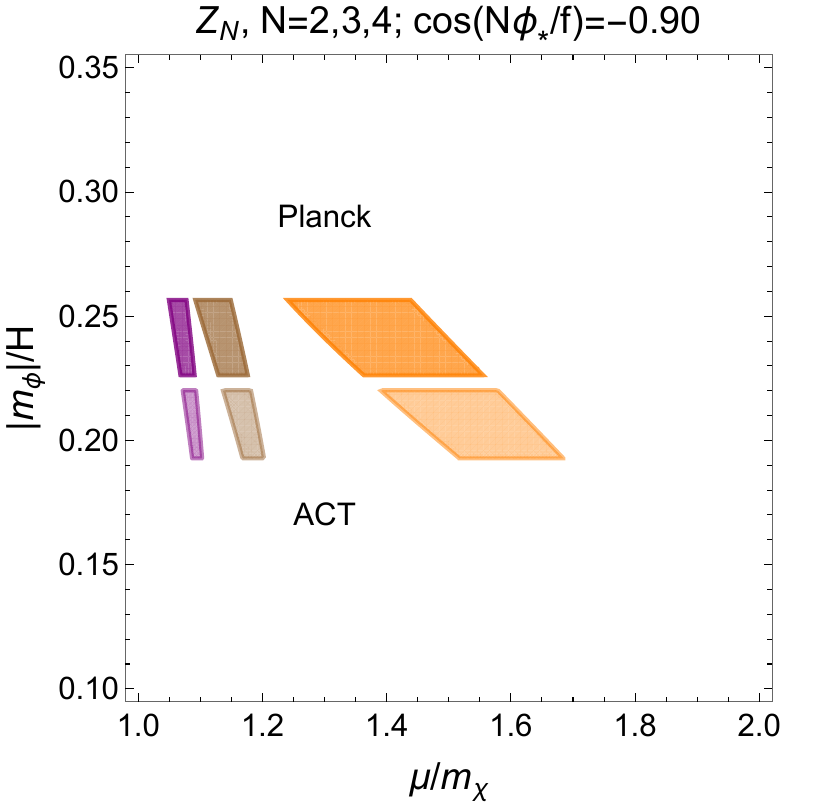} \,\, \includegraphics[width=0.45\textwidth]{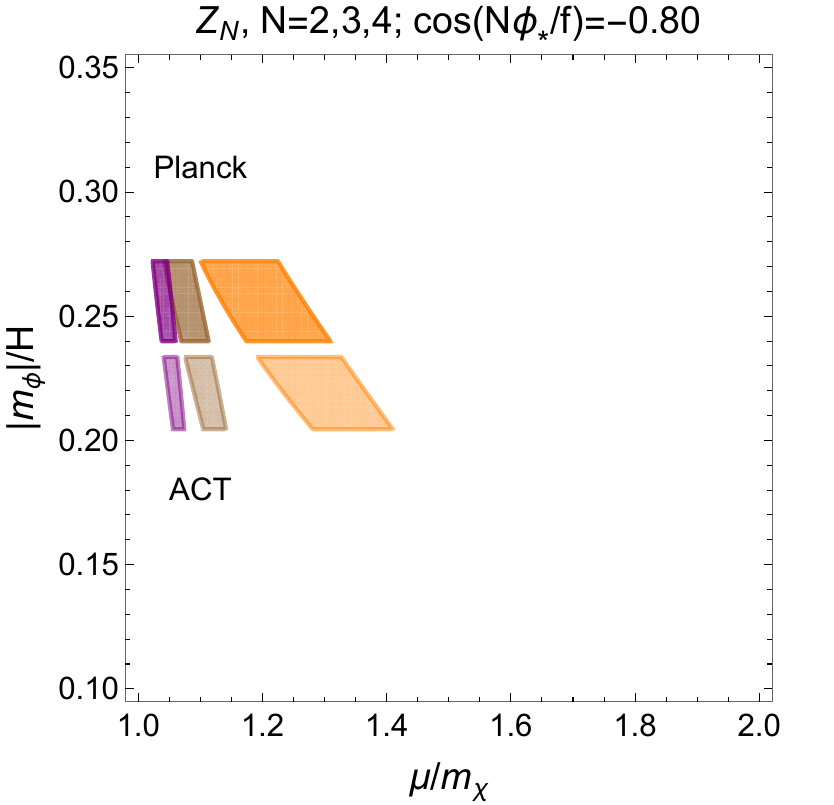}  \vspace{0.5cm} \\ 
\includegraphics[width=0.45\textwidth]{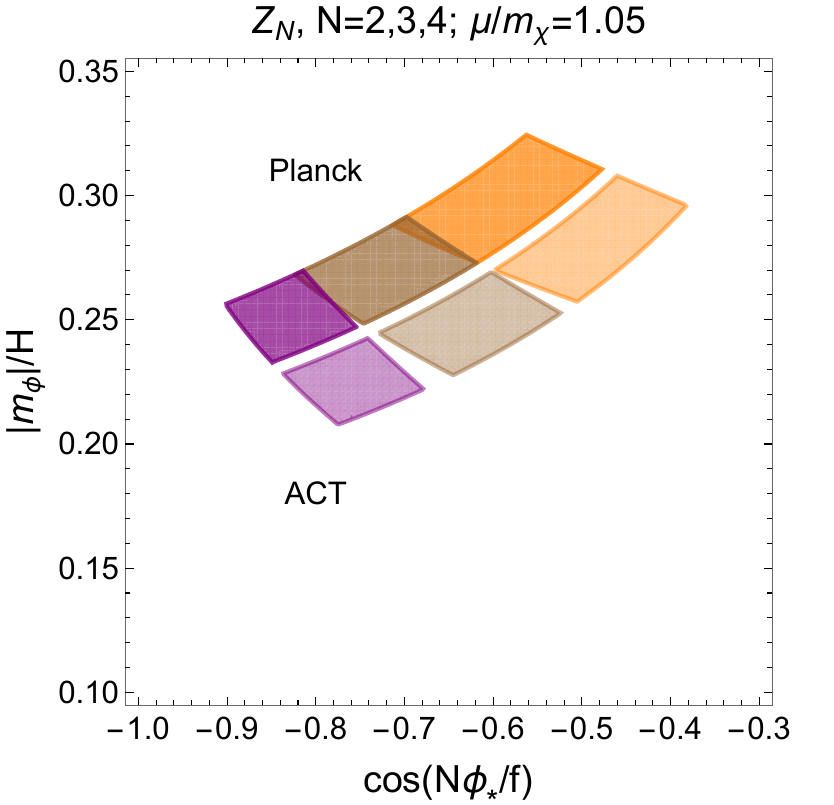} \,\, \includegraphics[width=0.45\textwidth]{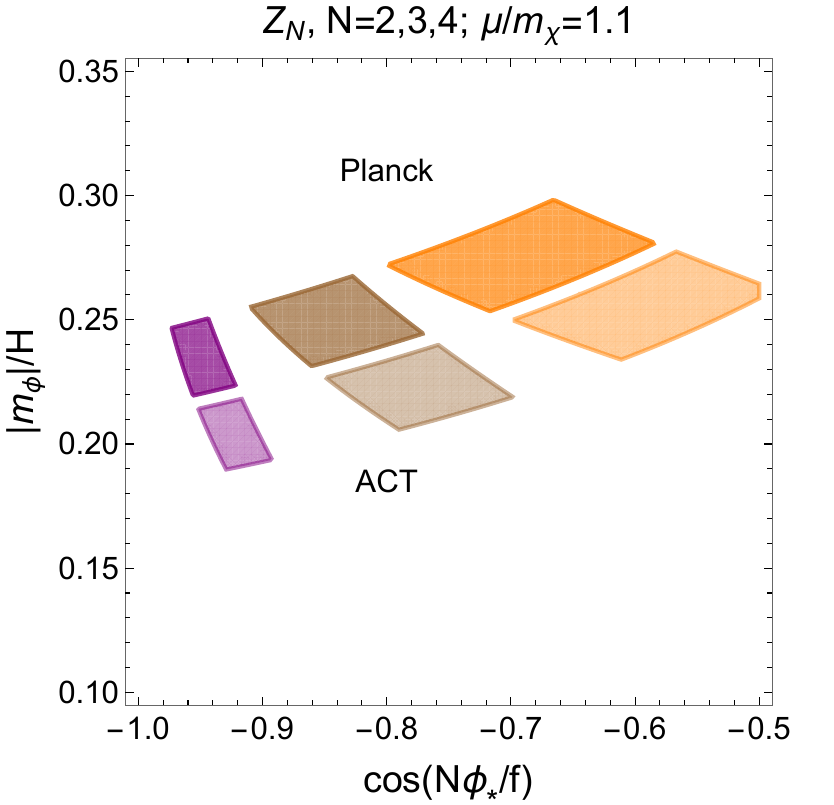} 
\caption{Parameter space for $Z_2,  Z_3, Z_4$ models at the one-loop order,  in orange, brown and purple regions, respectively, which is consistent with Planck data and $N=40-60$. Dark colors correspond to the Planck $1\sigma$ band and light colors correspond to the ACT $1\sigma$ band.}
\label{fig:zn}
\end{center}
\end{figure}

In Fig.~\ref{fig:zn}, we make a comparison between the inflationary predictions of the $Z_2,  Z_3, Z_4$ models with $N=40-60$ at one-loop, in orange, brown and purple regions, respectively,  showing the parameter space in each model that is consistent with Planck $1\sigma$ \cite{Planck} in thick colors and at ACT $1\sigma$ \cite{ACT} in light colors. In the upper panel, we show the parameter space for $a_0=\mu/m_\chi$ vs $|m^2_\phi|/H_I$, with a varying the inflation field value at the horizon exit, $\cos(N\phi_*/f)=-0.90, -0.80$ on left and right, respectively. On the other hand, in the lower panel, we also show the parameter space for $\cos(N\phi_*/f)$ vs $|m^2_\phi|/H_I$, with $\mu/m_\chi=1.05, 1.1$ on left and right, respectively.  

We find that for a fixed value of the inflaton at the horizon exit in the upper panel of Fig.~\ref{fig:zn}, the larger $N$, the closer $\mu/m_\chi$ gets to the narrow region around unity. This is because the range of the inflaton is restricted to $-\frac{\pi}{N}<\frac{\phi}{f}<\frac{\pi}{N}$ for the initial condition with $\frac{N\phi_*}{f}>-\pi$, so $\cos\big(\frac{\phi_c}{f}\big)$ gets close to unity for a large $N$, leading to $\mu^2/m^2_\chi=1/\cos\big(\frac{\phi_c}{f}\big)\gtrsim 1$ for the waterfall transition. On the other hand, for a fixed value of $\mu/m_\chi$ as in the lower panel of for a fixed value of $\mu/m_\chi$, the inflaton field value at the horizon exit is favored to be close to the top of the inflaton potential, as we go from $Z_2$ in orange to $Z_3$ in grown and to $Z_4$ in purple.

\section{Reheating and dark matter}

We discuss the vacuum structure of the inflaton and the waterfall fields with the $Z_N$ symmetry and show the UV insensitivity of the inflaton mass from loops with waterfall fields.  We also introduce the Higgs portal couplings for the waterfall fields and determine the reheating temperature from the perturbative decays of the waterfall field condensate. We also make a brief comment on the possibility of a multi-component dark matter from the waterfall fields.

\subsection{The vacuum structure}

From eq.~(\ref{fullpot}), the effective squared mass for the inflaton is given by
\bea
m^2_{\phi,{\rm eff}}=-\frac{N^2\Lambda^4}{f^2} \cos\bigg(\frac{N\langle\phi\rangle}{f}+\delta\bigg)+\frac{1}{2}\frac{\mu^2}{f^2}\sum_{k=1}^N \langle\chi^2_k\rangle \cos\bigg(\frac{\langle\phi\rangle}{f}+\frac{2\pi k}{N}\bigg)+m^2_{\phi,{\rm loops}}.
\label{effinfmass}
\eea
Here, the first term corresponds to a tachyonic mass for the inflaton during inflation and the second term comes from the inflaton couplings to the waterfall fields. The last term comes from the finite part of the CW potential, which corresponds to the loop corrections to $\eta V_{0,R}/M^2_P$. For instance,   for the $Z_2$ case, from eq.~(\ref{etaz20}), we obtain
\bea
m^2_{\phi,{\rm loops}} &\simeq& \frac{4{\Lambda}^4}{f^2} \,\frac{\mu^4}{128\pi^2{\Lambda}^4}  \bigg[\cos\Big(\frac{2\langle\phi\rangle}{f}\Big)\ln \bigg(\frac{m^4_\chi}{e M^2_{\chi,1} M^2_{\chi,2}}  \bigg) \nonumber \\
&&+1+\frac{m^2_\chi}{\mu^2} \cos\Big(\frac{\langle\phi\rangle}{f}\Big) \ln \bigg(\frac{M^2_{\chi,2}}{M^2_{\chi,1}}  \bigg) \bigg],
\eea  
with
\bea
M^2_{\chi,1}&=& m^2_\chi+3\lambda \langle\chi^2_1\rangle+\mu^2 \cos\Big(\frac{\langle\phi\rangle}{f}\Big), \\
M^2_{\chi,2}&=& m^2_\chi+3\lambda \langle\chi^2_2\rangle-\mu^2 \cos\Big(\frac{\langle\phi\rangle}{f}\Big)+\Delta_2.
\eea
Here, we note that $\Delta$ is the shifted waterfall field mass for the resummation given in eq.~(\ref{resum}) and the self-interactions of the waterfall fields are included. Then, at the end of inflation, the waterfall field $\chi_2$ takes a nonzero field value rapidly by the waterfall transition \cite{Garcia-Bellido:2001dqy,Lee:2023dcy}. Thus, the loop-corrected inflaton mass is well-defined from inflation all the way to the waterfall transition. Therefore, as we take $\mu^2\lesssim \Lambda^2$, in determining the vacuum structure, we can ignore the loop corrections to the inflaton mass and it is sufficient to take the tree-level potential in eq.~(\ref{fullpot}). 

Focusing on the tree-level contributions in eq.~(\ref{effinfmass}), at the end of inflation, the waterfall field with $m^2_{\chi_N}$ starts rolling fast at $\phi=\phi_c$, correcting the inflaton mass and stabilizing the inflaton at the common minimum for the inflaton potential and the waterfall-induced potential.

For simplicity, we take mixing masses and quartic couplings for waterfall fields to vanish \footnote{We note that it was shown in the $Z_2$ case that the general vacuum with $\alpha\neq 0$ and $\lambda'\neq 0$ can be stable \cite{Lee:2022fkd}, so we assume that a similar conclusion can be drawn for $N>2$ with nonzero mixing masses and quartic couplings.}, i.e. $\alpha=0$ and $\lambda'=0$, and assume that only the waterfall field $\chi_N$ takes $m^2_{\chi_N}<0$ during the waterfall transition. Then,  taking the phase in the inflaton potential to $\delta=\pi$, there is a stable minimum of the potential  at $\langle\phi\rangle=0$, $\langle\chi_N\rangle=v_\chi$, and $\langle\chi_j\rangle=0$  for $j=1,2,\cdots, N-1$, with
\bea
v_\chi = \sqrt{\frac{\mu^2-m^2_\chi}{\lambda}}. \label{vev}
\eea
Thus, the $Z_N$ symmetry is broken by the VEV of the waterfall field in the vacuum. The cosmological constant in the true vacuum can be fine-tuned to the observed value by
\bea
V_{\rm eff}(\chi_N=v_\chi, \chi_j=0)=V_0-\Lambda^4 -\frac{1}{4}\lambda \, v^4_\chi\simeq 0, \label{vacE}
\eea
so the vacuum energy dominated by $V_0$ during inflation is cancelled after the waterfall transition. 
As a result, combining eqs.~(\ref{vev}) and (\ref{vacE}), the quartic coupling $\lambda$ and the VEV $v_\chi$ in terms of the waterfall field masses and the Hubble scale during inflation as
\bea
\lambda=\frac{(\mu^2-m^2_\chi)^2}{12M^2_P H^2_I}, \qquad v_\chi=\frac{2\sqrt{3} M_P H_I}{\sqrt{\mu^2-m^2_\chi}}.
\eea

Now we expand the waterfall field around the vacuum by  $\chi_N=v_\chi+{\tilde \chi}_N$, we find that $(\phi, {\tilde \chi}_N, \chi_j)$ with $j=1,2,\cdots, N-1$, don't mix with one another, and their mass eigenvalues are
\bea
m^2_\phi&=&\frac{1}{f^2}\bigg(N^2\Lambda^4+\frac{1}{2}\mu^2 v^2_\chi \bigg), \\
m^2_{{\tilde \chi},N} &=& 2\lambda v^2_\chi \nonumber \\
&=& 2(\mu^2-m^2_\chi), \label{wmassvac} \\
m^2_{\chi,j} &=& m^2_\chi-\mu^2 \cos\Big(\frac{2\pi j}{N} \Big)+\lambda' v^2_\chi (\delta_{j,N-1}+\delta_{j,N+1}), \quad j=1,2,\cdots, N-1.
\eea

For instance, for the $Z_2$ case, we get
\bea
m^2_{\chi,1}&=&  m^2_\chi+\mu^2+ \lambda' v^2_\chi \nonumber \\
&=&m^2_\chi + \mu^2+\frac{\lambda'}{\lambda} (\mu^2-m^2_\chi).
\eea
Here, we note that $\lambda>0$ and $\lambda+\lambda'>0$ if $\lambda'<0$ from the vacuum stability. So, if $\mu^2/m^2_\chi>-3+\frac{8\lambda'}{\lambda}/(1+\frac{2\lambda'}{\lambda})$ for $\lambda'/\lambda>-\frac{1}{2}$ or $\mu^2/m^2_\chi>-3+\frac{8\lambda'}{\lambda}/(1+\frac{2\lambda'}{\lambda})$ for $\lambda'/\lambda<-\frac{1}{2}$, then $m_{{\tilde \chi},2}<2m_{\chi,1}$. In this case, the decay mode of the waterfall field, ${\tilde\chi}_2\to \chi_1\chi_1$, is not open in the vacuum.

For the $Z_3$ case, we get
\bea
m^2_{\chi,1}= m^2_\chi+\frac{1}{2}\mu^2+\lambda' v^2_\chi =m^2_{\chi,2}.
\eea
So, if $\mu^2/m^2_\chi<\frac{3}{2}\frac{\lambda}{\lambda'}-1$ for $\lambda'>0$ or $\mu^2/m^2_\chi>\frac{3}{2}\frac{\lambda}{\lambda'}-1$ for $\lambda'<0$, then $m_{{\tilde \chi},3}<2m_{\chi,1}, 2m_{\chi,2}$. In this case, the decay modes of the waterfall field, ${\tilde\chi}_3\to \chi_1\chi_1,\chi_2\chi_2$, are not open in the vacuum.

For the $Z_4$ case, we get
\bea
m^2_{\chi,1}&=& m^2_\chi+\lambda' v^2_\chi = m^2_{\chi,3}, \\
m^2_{\chi,2} &=& m^2_\chi + \mu^2.
\eea
So, $m_{{\tilde \chi},4}<2m_{\chi,2}$ is always true, so ${\tilde\chi}_4\to \chi_2\chi_2$ is kinematically forbidden.  If $\mu^2/m^2_\chi<3+\frac{4\lambda'}{\lambda}/(1-\frac{2\lambda'}{\lambda})$ for $\lambda'/\lambda<\frac{1}{2}$ or $\mu^2/m^2_\chi>3+\frac{4\lambda'}{\lambda}/(1-\frac{2\lambda'}{\lambda})$ for $\lambda'/\lambda>\frac{1}{2}$ , then $m_{{\tilde \chi},4} <2m_{\chi,1}=2m_{\chi,3}$. In this case, the decay modes of the waterfall field, ${\tilde\chi}_4\to \chi_1\chi_1,\chi_3\chi_3$, are not open in the vacuum.

\begin{figure}[!t]
\begin{center}
\includegraphics[width=0.45\textwidth]{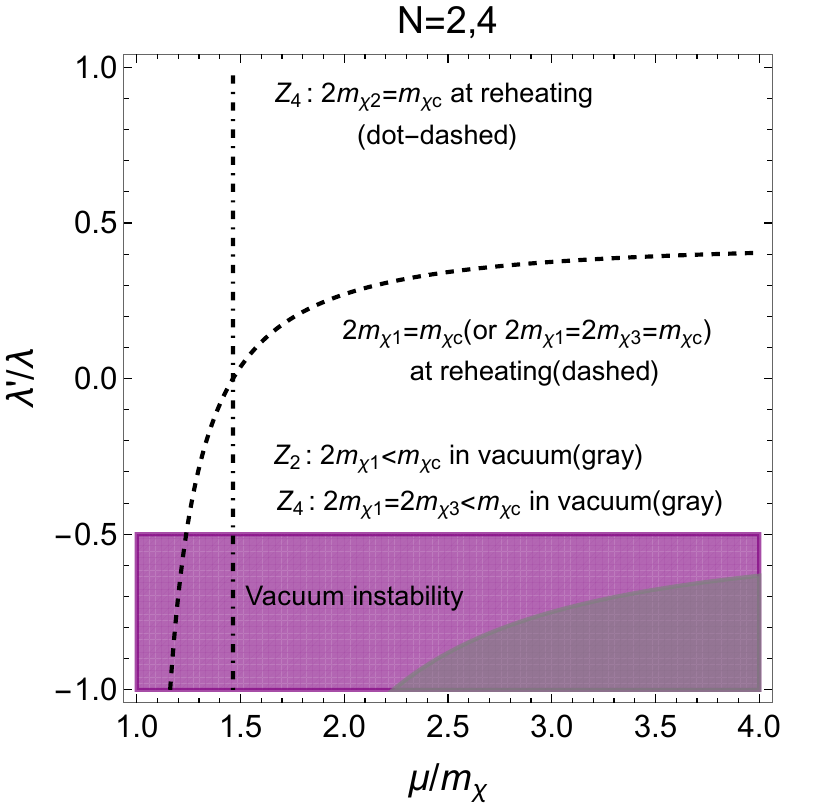} \,\, \includegraphics[width=0.45\textwidth]{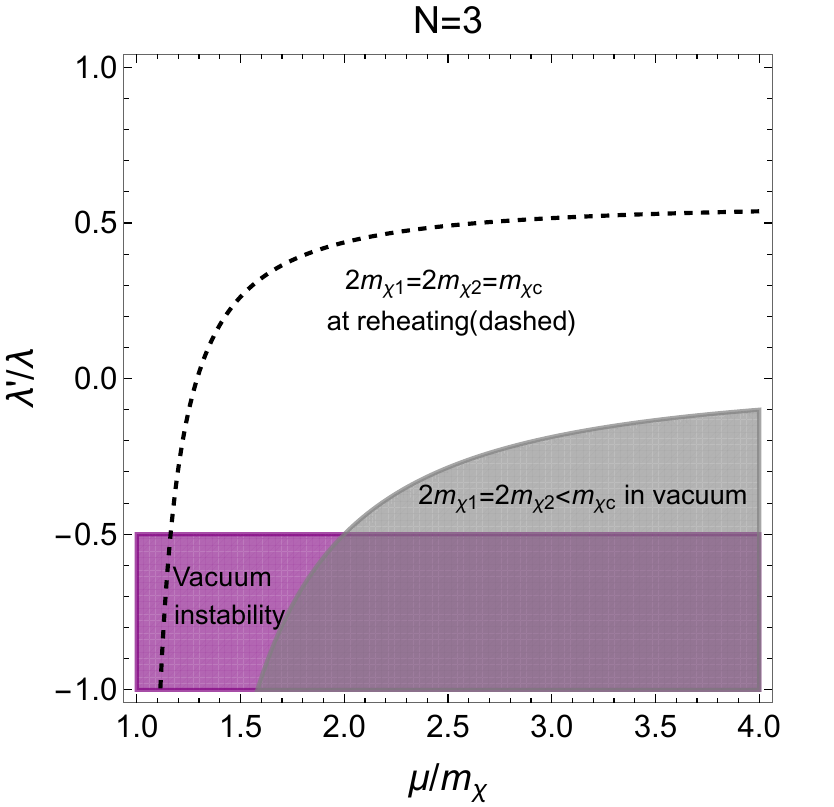}  
\caption{Parameter space for $\mu/m_\chi$ vs $\lambda'/\lambda$ in gray where the waterfall field condensate decays into a pair of other waterfall fields. In the gray regions, $2m_{\chi,1}<m_{\chi_c}\equiv m_{{\tilde \chi},2} $ for $Z_2$ and $2m_{\chi,1}=2m_{\chi,3}<m_{\chi_c}\equiv m_{{\tilde \chi},4} $ for $Z_4$ on left, and $2m_{\chi,1}=2m_{\chi,2}<m_{\chi_c}\equiv m_{{\tilde \chi},3} $ for $Z_3$ on right. For the $Z_4$ case, $2m_{\chi,2}>m_{\chi_c}$ in the vacuum, so the waterfall field condensate does not decay into a pair of $\chi_2\chi_2$. We also note that  the waterfall field condensate can decay into waterfall fields during reheating below the dashed lines in both plots and to the right of the dot-dashed line in the left plot. The absolute vacuum stability is not satisfied in the purple regions.}
\label{fig:wmass}
\end{center}
\end{figure}

In Fig.~\ref{fig:wmass}, we show the parameter space for $\mu/m_\chi$ vs $\lambda'/\lambda$ in gray where the waterfall field condensate in the vacuum decays into a pair of other waterfall fields. In the gray regions, the masses of the waterfall fields in the vacuum satisfy $2m_{\chi,1}<m_{\chi_c}\equiv m_{{\tilde \chi},2} $ for $Z_2$ and $2m_{\chi,1}=2m_{\chi,3}<m_{\chi_c}\equiv m_{{\tilde \chi},4} $ for $Z_4$ on left, and $2m_{\chi,1}=2m_{\chi,2}<m_{\chi_c}\equiv m_{{\tilde \chi},3} $ for $Z_3$ on right. As we noted, $2m_{\chi,2}>m_{\chi_c}$ in the vacuum for the $Z_4$ case, so the waterfall field condensate does not into a pair of $\chi_2\chi_2$ in this case.  As a result, the decays of the waterfall field condensate into other waterfall fields require relatively large couplings for the waterfall field condensate, $\mu/m_\chi$ and $\lambda'/\lambda$,  in the $Z_2$ and $Z_4$ cases, as compared to the $Z_3$ case. But, relatively large waterfall couplings to the inflaton in the gray regions in Fig.~\ref{fig:wmass} are disfavored for the successful inflation and some parameter space with a negative $\lambda'$ is not consistent with the absolute vacuum stability as shown in the purple regions.  Nonetheless, the effective mass of the waterfall field condensate becomes larger during reheating than the one in the vacuum, decaying into other waterfall fields in the regions below the dashed lines and to the right of the dot-dashed line in Fig.~\ref{fig:wmass}.

We remark that the cancellation of quadratic divergences for the inflaton mass still occurs in the true vacuum as during inflation, independent of  the order of the $Z_N$ symmetry. 
We first identify the interaction terms up to quadratic orders in each of  $(\phi, {\tilde \chi}_N, \chi_j)$, with $j=1,2,\cdots, N-1$, as follows,
\bea
{\cal L}_{{\rm int, eff}} &=&- \frac{1}{2}\mu^2 v_\chi {\tilde\chi}_N \bigg(\frac{\phi}{f}\bigg)^2 -\frac{1}{4}\mu^2 {\tilde\chi}^2_N \bigg(\frac{\phi}{f}\bigg)^2 \nonumber \\
&&+\frac{1}{2}  \mu^2\bigg(\frac{\phi}{f}\bigg)\sum_{k=1}^{N-1} \chi^2_k \sin\Big(\frac{2\pi k}{N} \Big) -\frac{1}{4}\mu^2  \bigg(\frac{\phi}{f}\bigg)^2\,\sum_{k=1}^{N-1} \chi^2_k \cos\Big(\frac{2\pi k}{N} \Big).
\eea
As a result, we first find that $\phi^2{\tilde\chi}_N$ and $\phi\chi^2_j$ couplings have a mass dimension-one, so they do not give rise to quadratic divergences for the inflaton mass. Secondly, $\phi^2{\tilde\chi}^2_N$ and $\phi^2\chi^2_j$ couplings lead quadratically divergent contributions to the inflaton mass individually, but those are summed up to zero due to  the $Z_N$ symmetry. The reason is the following. The corrections to the inflaton mass due to $\phi^2{\tilde\chi}^2_N$ and $\phi^2\chi^2_j$ are
\bea
\Delta m^2_\phi&=&-i\frac{\mu^2}{4f^2} \int \frac{d^4 k}{(2\pi)^4}\bigg(\frac{1}{k^2-m^2_{{\tilde\chi}_N}}+\sum_{j=1}^{N-1} \frac{\cos\big(\frac{2\pi j}{N}\big)}{k^2-m^2_{\chi_j}} \bigg) \nonumber \\
&=&-i\frac{\mu^2}{4f^2} \int \frac{d^4 k}{(2\pi)^4}\bigg(\frac{1}{k^2-m^2_{{\tilde\chi}_N}}+\sum_{j=1}^{N-1}\frac{\cos\big(\frac{2\pi j}{N}\big)}{k^2-m^2_{\chi_1}}\bigg) \nonumber \\
&&-i\frac{\mu^2}{4f^2}\sum_{j=1}^{N-1} \cos\Big(\frac{2\pi j}{N}\Big) \int \frac{d^4 k}{(2\pi)^4}\bigg( \frac{1}{k^2-m^2_{\chi_j}}- \frac{1}{k^2-m^2_{\chi_1}} \bigg) \nonumber \\
&=&-i\frac{\mu^2}{4f^2}\,(m^2_{{\tilde \chi}_N}-m^2_{\chi_1}) \int \frac{d^4 k}{(2\pi)^4}\frac{1}{(k^2-m^2_{{\tilde\chi}_N})(k^2-m^2_{\chi_1})} \nonumber \\
&&-i\frac{\mu^2}{4f^2}\sum_{j=1}^{N-1} \cos\Big(\frac{2\pi j}{N}\Big) (m^2_{\chi_j}-m^2_{\chi_1})\int \frac{d^4 k}{(2\pi)^4}\frac{1}{(k^2-m^2_{\chi_j})(k^2-m^2_{\chi_1})}. \label{inflatonmass}
\eea
Here, we used $\sum_{k=1}^N \cos\big(\frac{2\pi k}{N}\big)=0$ in the third line. 
Therefore, each term in the last line in eq.~(\ref{inflatonmass}) shows the cancellations of the quadratic divergences, leaving only the logarithmic divergences, namely,
\bea
\Delta m^2_\phi &=&\frac{1}{(4\pi)^2}\,\frac{\mu^2}{4f^2}  \bigg[(m^2_{{\tilde\chi}_N}-m^2_{\chi_1}) +\sum_{j=1}^{N-1} \cos\Big(\frac{2\pi j}{N}\Big) \Big(m^2_{\chi_j} -m^2_{\chi_1}\Big)\bigg] \ln \Big(\frac{M^2_*}{m^2_\chi}\Big) \nonumber \\
&=&\frac{1}{(4\pi)^2}\,\frac{\mu^2}{4f^2} \sum_{j=1}^{N-1} \cos\Big(\frac{2\pi j}{N}\Big) \Big(m^2_{\chi_j} -m^2_{{\tilde\chi}_N}\Big) \ln \Big(\frac{M^2_*}{m^2_\chi}\Big).
\eea

\subsection{Reheating}

In this subsection, we discuss the reheating after the waterfall transition in the $Z_N$ models, generalizing the $Z_2$ case in Ref.~\cite{Lee:2023dcy}.

We consider the $Z_N$ invariant Higgs-portal couplings of the waterfall fields to the SM Higgs, as follows,
\bea
{\cal L}_{\rm H-portal}=-\kappa_1|H|^2 \sum_{k=1}^N \chi^2_k -\kappa_2 |H|^2 \sum_{k=1}^N \chi_k \chi_{k+1}.
\eea

Then, taking the waterfall fields by $\chi_N(t)=v_\chi+\chi_c(t)$ and $\chi_j(t)=0$  with  $j=1,2,\cdots, N-1$ during reheating and including the mixing quartic coupling $\lambda'$ for the waterfall fields, we obtain the effective interactions for the SM Higgs as
\bea
{\cal L}_{H,{\rm eff}}&=& -\kappa_1|H|^2(2v_\chi \chi_c+\chi^2_c)-\kappa_2|H|^2 (v_\chi+\chi_c)(\chi_{N-1}+\chi_1) \nonumber \\
&& -\frac{1}{2}\lambda' (2v_\chi \chi_c +\chi^2_c)(\chi^2_{N-1}+\chi^2_1).
\eea
Here, we note that the effective masses for the SM Higgs and the waterfall fields other than $\chi_N$ are given by
\bea
m^2_{H,{\rm eff}} &=& m^2_{H,0}+ \kappa_1 \chi^2_N(t), \\
m^2_{\chi, j, {\rm eff}} &=& m^2_{\chi,j,0}+\lambda' \chi^2_N(t) (\delta_{j,N-1}+\delta_{j,N+1}), \quad j=1,2,\cdots, N-1,
\eea
where $m^2_{H,0}$ and $m^2_{\chi, j, 0}=m^2_\chi+m^2_j$ are the squared bare masses, being independent of the field value of $\chi_N$.
We note that the effective mass of the waterfall field condensate $\chi_c$ is given by 
\bea
m^2_{\chi_c} =\frac{\partial^2 V}{\partial \chi^2_N}=m^2_{\chi,N}+ 3\lambda \chi^2_N(t),
\eea
with $m^2_{\chi,N}=m^2_\chi-m^2_N$. We note that for the initial amplitude $v_\chi$ of the waterfall field condensate $\chi_c(t)$, the value of the effective mass of the waterfall field condensate gets larger as $m^2_{\chi_c}\leq 11(\mu^2-m^2_\chi)=\frac{11}{2} m^2_{{\tilde\chi},N}$, as compared to the waterfall field mass in the vacuum in eq.~(\ref{wmassvac}). Then, the waterfall field condensate can decay into $\chi_1$ or $\chi_{N-1}$ during the oscillation of the waterfall condensate.

As a result, we can determine the reheating temperature by the perturbative decay of the waterfall field condensate into an SM Higgs pair or a pair of the other waterfall fields by
\bea
T_{\rm RH}=\bigg(\frac{90}{\pi^2 g_{\rm RH}}\bigg)^{1/4} \sqrt{M_P \Gamma_{\chi_c}}
\eea
where $g_{\rm RH}$ is the number of relativistic degrees of freedom at reheating completion and $\Gamma_{\chi_c}$ is the decay rate of the waterfall field condensate $\chi_c$, given by
\bea
\Gamma_{\chi_c}=\Gamma_{\chi_c\to H{\bar H}} + \Gamma_{\chi_c\to \chi_1\chi_1}+ \Gamma_{\chi_c\to \chi_{N-1}\chi_{N-1}},
\eea 
with
\bea
\Gamma_{\chi_c\to H{\bar H}} &=& \frac{\kappa_1^2 v^2_\chi}{4\pi m_{\chi_c}}\sqrt{1-\frac{4m^2_H}{m^2_{\chi_c}}}, \\
\Gamma_{\chi_c\to \chi_1\chi_1} &=&\frac{\lambda^{\prime 2} v^2_\chi}{4\pi m_{\chi_c}}   \sqrt{1-\frac{4m^2_{\chi,1}}{m^2_{\chi_c}}},  \\
\Gamma_{\chi_c\to \chi_{N-1}\chi_{N-1}}&=&\frac{\lambda^{\prime 2} v^2_\chi}{4\pi m_{\chi_c}}   \sqrt{1-\frac{4m^2_{\chi,N-1}}{m^2_{\chi_c}}}.
\eea
Then, if $\kappa_1\gtrsim \lambda'$, $\chi_c\to H{\bar H}$ is a dominant channel for reheating, so the reheating temperature is determined to be
\bea
T_{\rm RH}=\bigg(\frac{90}{\pi^2 g_{\rm RH}}\bigg)^{1/4} \bigg(\frac{\kappa_1^2}{8\pi \lambda}\bigg)^{1/2}\sqrt{M_P m_{\chi_c}}.
\eea
Then,  we can rewrite the reheating temperature as
\bea
T_{\rm RH}=0.11m_{\chi_c} \bigg(\frac{100}{g_{\rm RH}}\bigg)^{1/4} \bigg(\frac{\kappa_1^2}{ \lambda}\bigg)^{1/2} \bigg(\frac{M_P}{m_{\chi_c}}\bigg)^{1/2}. \label{RH}
\eea
For $\kappa^2_1/\lambda\lesssim 83 (m_{\chi_c}/M_P)$, we find that the reheating temperature is smaller than the tachyonic mass of the waterfall field $\chi_c$, so the $Z_N$ symmetry would not be restored during reheating and there is no domain wall problem associated with the $Z_N$ symmetry after reheating \cite{Lee:2023dcy}. For instance, for $H_I/f\sim 10^{-6}$ from eqs.~(\ref{Hubblez2}) or (\ref{Hubblezn})  and $f=10^9-10^{16}\,{\rm GeV}$, the Hubble scale during inflation is in the range of $H_I\sim 10^{3}-10^{10}\,{\rm GeV}$. Then, the reheating temperature is favored to be $T_{\rm RH}\lesssim m_{\chi_c}\sim 100 H_I\sim 10^{5}-10^{12}\,{\rm GeV}$.

We also note that for $\alpha=0$, there are separate $Z_2$ symmetries for the waterfall fields, which are broken spontaneously during the waterfall transition. In our examples, the waterfall transition takes place along the $\chi_N$ direction, leading to a nonzero VEV for $\chi_N$, while the other waterfall fields do not get VEVs. We can always choose the inflaton direction such that  the universe evolves only to the vacuum with $\langle\chi_N\rangle\neq 0$ through the waterfall transition as discussed above. Thus, the $Z_2$ domain walls dividing $\langle\chi_N\rangle=+v_\chi$ from $\langle\chi_N\rangle=-v_\chi$ can be formed after the waterfall transition. In this case, in order to avoid a domain-wall problem, we need to introduce a small bias term at least for $\chi_N$ such that the formed domain-walls become unstable. However, we can still keep separate $Z_2$ symmetries for the rest of the waterfall fields without domain-wall problems, as far as  the reheating temperature satisfies $T_{\rm RH}\lesssim m_{\chi_c}$.

We remark that the maximum temperature of the universe can be higher than the reheating temperature \cite{maxtemp,Lee:2023dcy}. So, when the waterfall field is matter-like, the maximum reheating temperature is given \cite{maxtemp,Lee:2023dcy} by 
\bea
T_{\rm max}=0.77\, T_{\rm RH} \bigg(\frac{72}{5\pi^2 g_{\rm RH}}\bigg)^{1/8} \bigg(\frac{H_I M_P}{T^2_{\rm RH}}\bigg)^{1/4}. \label{Tmax}
\eea
For instance, for $T_{\rm RH}\lesssim 10^{5}\,{\rm GeV}-10^{12}\,{\rm GeV}$ and $m_{\chi_c}\sim 100 H_I$, it is sufficient to take $H_I\sim 10^{3}\,{\rm GeV}-10^{10}\,{\rm GeV}$ for $m_{\chi_c}\gtrsim T_{\rm RH}$. However, in this case, using eq.~(\ref{Tmax}) with $g_{\rm RH}=106.75$, we obtain the maximum temperature as $T_{\rm max}\lesssim 3.2\times 10^7\,{\rm GeV}-5.6\times 10^{12}\,{\rm GeV}$, which is larger than $m_{\chi_c}$, so the $Z_N$ symmetry could be restored before reheating completion. Therefore, we need to take a large waterfall mass by $m_{\chi_c}\gtrsim (560-3.2\times 10^4)H_I$ for $H_I\sim 10^{3}\,{\rm GeV}-10^{10}\,{\rm GeV}$ or introduce a bias term breaking the $Z_N$ symmetry explicitly.

In the $Z_2$ model, it was shown that the tachyonic instability after inflation can become important for reheating \cite{Garcia-Bellido:2001dqy} as well as dark matter production, provided that $\kappa_1/\lambda\lesssim 2.6\times 10^{-5}\sqrt{100H_I/m_1}$ \cite{Lee:2023dcy}. A similar conclusion can be drawn for the $Z_N$ cases with $N>2$.
We assume that preheating \cite{Lee:2023dcy} is not efficient and the reheating process is not instantaneous.
Then, the number of efoldings required to solve the horizon problem \cite{reheating,HiggsR2} is modified to
\bea
N_e=61.1 +\Delta N_e -\ln \bigg(\frac{V^{1/4}_0}{H_k} \bigg) -\frac{1}{12} \ln \bigg(\frac{g_{\rm RH}}{106.75} \bigg)
\eea
where the effects due to to the non-instantaneous reheating are encoded into
\bea
\Delta N_e= \frac{1}{12} \bigg(\frac{3w-1}{w+1} \bigg) \ln\bigg(\frac{45 \rho_{\chi_N}(t_c)}{\pi^2 g_{\rm RH} T^4_{\rm RH}} \bigg).\label{DN}
\eea
Here, $\rho_{\chi_N}(t_c)$ is the energy density of the waterfall field at the end of the waterfall transition at $t_c$ and $w$ is the equation of state for the waterfall condensate.
Here, $H_k$ is the Hubble parameter evaluated at the horizon exit for the Planck pivot scale, $k=0.05\,{\rm Mpc}^{-1}$, and $w$ is the averaged equation of state during reheating. 
Then, taking $w=0$ for the matter-like waterfall field condensate and $g_{\rm RH}=106.75$ in eq.~(\ref{DN}), we obtain the number of efoldings  as
\bea
N_e=51.3-\frac{1}{3}\ln\bigg(\frac{H_I}{1.6\times 10^{10}\,{\rm GeV}}\bigg)+\frac{1}{3}\ln\bigg(\frac{T_{\rm RH}}{10^{14}\,{\rm GeV}}\bigg).
\eea
As a result,  the number of efoldings varies depending on the Hubble scale during inflation and the reheating temperature, determined by eq.~(\ref{RH}). If the reheating temperature is low such that $T_{\rm RH}\lesssim H_I \lesssim m_{\chi_c}$,  there is no restoration of the $Z_N$ symmetry after reheating. For instance, for $T_{\rm RH}\simeq H_I$,  the number of efoldings becomes $N_e\simeq 48$. For a high reheating temperature satisfying $T_{\rm RH}\gg m_{\chi_c}\gtrsim H_I$, the number of efoldings gets larger. For instance, $T_{\rm RH}\simeq 10^3\, m_{\chi_c}$, the number of efoldings becomes $N_e\simeq 51$, but the $Z_N$ symmetry would be restored so we would need to introduce small terms breaking the $Z_N$ symmetry explicitly to destabilize the domain-walls.

\subsection{Multi-component dark matter from waterfall fields}

For $\alpha=0$ and $\kappa_2=0$, there is an accidental $Z'_2$ symmetry, under $\chi_j\to- \chi_j$, with $j=1,2,\cdots, N-1$, except $\chi_c$ which gets a nonzero VEV. Thus, the waterfall fields, $\chi_j$'s, can be candidates for dark matter.  Due to the $Z'_2$ invariant Higgs-portal couplings, $\kappa_1$, and/or the couplings of the waterfall field condensate $\chi_c$ to $\chi_j$'s,  we can get $N-1$ components of dark matter out of the waterfall fields, $\chi_j$'s. As $\chi_1$ and $\chi_{N-1}$ couple to the waterfall field condensate $\chi_c$ directly, they can be produced from the decays or scatterings of $\chi_c$, if kinetically allowed, as illustrated in the gray regions in Fig.~\ref{fig:wmass}.  On the other hand, the rest of $\chi_j$'s can be produced from the loop decays of $\chi_N$, and through the scatterings of $\chi_1\chi_1\to \chi_2\chi_2$,  $\chi_{N-1}\chi_{N-1}\to \chi_{N-2}\chi_{N-2}$, etc, or  $H {\bar H}\to \chi_j \chi_j$.

The freeze-in production of the waterfall field dark matter during and after reheating was discussed for the $Z_2$ case \cite{Lee:2023dcy}. Waterfall fields $\chi_j$'s have comparable masses to the one for $\chi_c$, namely, $m_{\chi,j}\sim m_{\chi_c}$, so heavy dark matter should be produced non-thermally from the decays of $\chi_c$ if $m_{\chi_c}\sim H_I\gtrsim 100\,{\rm TeV}$ (i.e. the unitarity limit for WMP dark matter) during inflation. 
A similar discussion in the $Z_2$ case can be applied to  $\chi_j$'s in the $Z_N$ case, but we don't pursue the analysis the dark matter phenomenology further in this work.

\section{Conclusions}

We presented a new model for pseudo-Nambu-Goldstone inflation in the presence of $N$ waterfall scalar fields. 
The $Z_N$ symmetry sets the couplings  between the inflaton and the waterfall fields such that the stability of the slow-roll inflation is ensured under the loop corrections and  inflation ends due to the waterfall transition along one of the waterfall fields. 

Considering the Coleman-Weinberg potential for the inflaton during inflation, we showed that the quadratically divergent loop corrections are cancelled between the waterfall fields, independent of $N$, and even the logarithmically divergent loop corrections are absent for $N>2$, making the inflaton potential insensitive to the UV physics. First, in the $Z_2$ case, a successful inflation is maintained under the logarithmic loop corrections as far as smaller waterfall couplings or larger inflaton masses are chosen.
For higher order $Z_N$ symmetries with $N>2$,  the field excursion of the inflaton during inflation gets shrunken to $-\frac{\pi}{N}<\frac{\phi}{f}<\frac{\pi}{N}$, so the waterfall field transition occurs close to the top of the inflaton potential or the waterfall mass parameters are constrained to $\mu^2/m^2_\chi=1/\cos\big(\frac{\phi_c}{f}\big)\gtrsim 1$.

We also discussed the implications of the inflationary conditions on the waterfall mass parameters for the vacuum structure and the reheating. Reheating in the SM sector occurs dominantly through the $Z_N$-invariant Higgs-portal couplings. 
We showed that the maximum temperature can be lower than the mass of the waterfall field condensate such that the $Z_N$ symmetry is not restored after reheating, so there is no domain wall problem in this case.
During reheating, we showed that the waterfall field condensate can decay into the neighboring waterfall fields, $\chi_1$ and $\chi_{N-1}$, directly, annihilating into other waterfall fields, so there is a possibility of multi-component dark matter from the waterfall fields in the presence of accidental $Z'_2$ symmetries, namely, when the mass mixings for the waterfall fields vanish.

\section*{Acknowledgements}

HML is supported in part by Basic Science Research Program through the National
Research Foundation of Korea (NRF) funded by the Ministry of Education, Science and
Technology (NRF-2022R1A2C2003567). 
AM acknowledges support by the Deutsche Forschungsgemeinschaft (DFG, German Research Foundation) under the DFG Emmy Noether Grant No. PI 1933/1-1 and Germany’s Excellence Strategy – EXC 2121 “Quantum Universe” – 390833306.

\appendix 
\section{Effects of mixing quartic couplings}\label{mixquartic}

In the Appendix, we show the effects of mixing quartic couplings for waterfall fields on the tree-level effective potential for the inflaton.

\underline{\bf The tree-level effective potential for the inflaton}

Setting the mixing mass terms for waterfall fields, namely, $\alpha=0$, we can write the waterfall field potential in matrix notations,
\bea
V_W=(M^2)^T \chi^2 + (\chi^2)^T \Lambda \chi^2
\eea
where $(M^2)^T=(m^2_{\chi,1},m^2_{\chi,2},\cdots,m^2_{\chi,N})$ with $m^2_{\chi,k}=m^2_\chi+m^2_k$, $(\chi^2)^T=(\chi^2_1,\chi^2_2,\cdots,\chi^2_N)$ and $\Lambda$ is the $N\times N$ real symmetric matrix, given by 
\bea
\Lambda =\left( \begin{array}{ccccccccc}  \lambda & \lambda' & 0 & 0 & 0  &   \cdots &  0 & \lambda'  \\  \lambda' & \lambda  & \lambda'  & 0  & 0  & \cdots & 0 &  0  \\ 0 & \lambda' & \lambda  & \lambda' & 0  &  \cdots & 0 &  0  \\  0 & 0  & \lambda' & \lambda  & \lambda'  & \cdots & 0  & 0   \\  0 & 0 &  0 &  \lambda' & \lambda & \cdots & 0 & 0  \\  \vdots & \vdots &\vdots  & \vdots  & \vdots & \ddots  &  \vdots & \vdots \\ 0 & 0 & 0 & 0 & 0 &   \cdots &  \lambda & \lambda'
\\  \lambda' & 0 & 0 & 0 & 0 &   \cdots & \lambda' & \lambda  \end{array} \right).
\eea
Then, the minimization conditions, $\frac{\partial V_W}{\partial \chi_k}=0$, for nonzero waterfall field VEVs, are given by
\bea
(\langle\chi^2\rangle)^T =-\frac{1}{2} (M^2)^T \Lambda^{-1} \label{solutions}
\eea
or
\bea
\langle\chi^2\rangle =- \frac{1}{2} \Lambda^{-1} M^2.
\eea
Therefore, if all the waterfall fields get nonzero VEVs, we integrate out the waterfall fields to get the effective potential for $\phi$ as follows,
\bea
V_{1,\rm eff}= -\frac{1}{4} (M^2)^T \Lambda^{-1} M^2.  \label{effpot}
\eea

Suppose that some of waterfall fields VEVs vanish, i.e., $\langle \chi_a\rangle=0$, for $a=N-M+1, N-M+2,\cdots, N$. In this case, only the waterfall fields with nonzero VEVs contribute to the effective potential, leading to 
\bea
V_{1,\rm eff}&=& -\frac{1}{4} \sum_{i,j\neq a} ((M^2)^T\Lambda^{-1})_i \Lambda_{ij} (\Lambda^{-1} M^2)_j \nonumber \\
&=& -\frac{1}{4} (M^2)^T_j (\Lambda^{-1})_{jk} (M^2)_k + \frac{1}{4} ((M^2)^T \Lambda^{-1})_a \Lambda_{ab} (\Lambda^{-1} M^2)_b \nonumber \\
&=&\frac{1}{4} ((M^2)^T \Lambda^{-1})_a \Lambda_{ab} (\Lambda^{-1} M^2)_b +{\rm constant}.   \label{effpot2}
\eea
Here, we used the fact that the full sum in eq.~(\ref{effpot}) is constant.
\vspace{0.5cm}

\noindent
\underline{\bf The case with $Z_3$ waterfalls}

For instance, for $N=3$, the quartic coupling matrix becomes
\bea
\Lambda= \left( \begin{array}{ccc} \lambda & \lambda' & \lambda' \\  \lambda' & \lambda & \lambda' \\  \lambda' & \lambda' & \lambda \end{array} \right).
\eea
In this case, the inverse matrix $\Lambda^{-1}$ can be obtained to be
\bea
\Lambda^{-1}= \frac{\kappa}{3} \left( \begin{array}{ccc} 1 & 1 & 1 \\  1 & 1 & 1  \\  1 & 1 & 1 \end{array} \right)+ \frac{\kappa'}{3} \left( \begin{array}{ccc} 2 & -1 & -1 \\  -1  & 2 & -1 \\  -1 & -1& 2 \end{array} \right)
\eea
with
\bea
\kappa=\frac{1}{\lambda+2\lambda'}, \quad  \kappa'=\frac{1}{\lambda-\lambda'}. 
\eea
Then,  we get
\bea
\langle\chi^2\rangle=-\Lambda^{-1} M^2= -\left( \begin{array}{c} \kappa m^2_\chi + \frac{\kappa'}{3}(2m_1^2-m^2_2-m^2_3) \\ 
\kappa m^2_\chi + \frac{\kappa'}{3}(-m_1^2+2m^2_2-m^2_3)
\\  \kappa m^2_\chi + \frac{\kappa'}{3}(-m_1^2-m^2_2+2m^2_3) \end{array} \right) = -\left( \begin{array}{c} \kappa m^2_\chi + \kappa' m_1^2 \\ 
\kappa m^2_\chi +  \kappa' m_2^2 
\\  \kappa m^2_\chi + \kappa' m_3^2 \end{array} \right)
\eea
where use is made of $m_1^2+m_2^2+m_3^2=0$ in the second equality. 
Here, we note that symmetry breaking conditions are $\kappa m^2_\chi+\kappa' m^2_k<0$, which are different from $m^2_{\chi_k}=m^2_\chi+m^2_k<0$ with no mixing quartic couplings, namely, $\lambda'=0$.

First, in the configuration where the VEVs of all the waterfall fields are nonzero, the effective potential (\ref{effpot}) would become
\bea
V_{1,\rm eff}(\phi)&=& -\frac{1}{4} \sum_{k=1}^3 (m^2_\chi+m^2_k) (\kappa m^2_\chi +\kappa' m^2_k) \nonumber \\
&=& -\frac{3}{4} \kappa m^4_\chi+ \frac{1}{4} \sum_{k=1}^3  \Big( -(\kappa+\kappa')m^2_\chi m^2_k +\kappa' m^4_k \Big) \nonumber \\
&=&  -\frac{3}{4} \kappa m^4_\chi+ \frac{3}{8}\,\kappa' \mu^4 = {\rm constant}. 
\eea
In the case with $\lambda'\neq 0$, we can regard $a_0=\frac{m^2_\chi}{\mu^2}$ as being replaced by $a'_0=\frac{\kappa m^2_\chi}{\kappa'\mu^2}$.  However, for $\kappa,\kappa'>0$ and $a'_0<1$, there is no field space satisfying $\kappa m^2_\chi + \kappa' m_i^2<0$ with $i=1,2,3$ simultaneously, so  there is no configuration where  the VEVs of all the waterfall fields are nonzero, as in the case with $\lambda'=0$.

When one of the waterfall fields VEVs vanish in Configuration II, for instance, $\langle \chi_3\rangle=0$,  from eq.~(\ref{effpot2}), we get the effective potential as
\bea
V_{1,\rm eff}(\phi)&=& \frac{1}{4} ((M^2)^T \Lambda^{-1})_3 \Lambda_{33} (\Lambda^{-1}M^2)_3 +{\rm const}  \nonumber \\
&=& \frac{\lambda}{4} (\kappa m^2_\chi +\kappa' m^2_3)^2+{\rm constant}  \nonumber \\
&=&\frac{1}{4} \lambda\kappa^{\prime 2} \mu^4 \Big( \cos \Big(\frac{\phi}{f}\Big)-b_0\Big)^2 +{\rm constant}  
\eea
with
\bea
b_0\equiv \frac{\kappa}{\kappa'}\, \frac{m^2_\chi}{\mu^2}. 
\eea
Thus, the parameters of the effective potential depends on the waterfall mixing quartic couplings.
\vspace{0.5cm}

\noindent
\underline{\bf The case with $Z_N$ waterfall fields}

We present  the tree-level effective potential for the inflaton from $N$ waterfall fields with vanishing mixing masses. 

In this case, we first get the inverse matrix $\Lambda^{-1}$ as 
\bea
\Lambda^{-1}= U^\dagger (\Lambda^{\rm diag})^{-1} U
\eea
where
\bea
 (\Lambda^{\rm diag})^{-1}= {\rm diag}(\kappa_1,\kappa_2,\cdots,\kappa_N)
\eea
with $1/\kappa_k=\lambda+2\lambda' \cos\Big(\frac{2(k-1) \pi}{N} \Big)$,
and 
\bea
U_{lm}=\frac{1}{\sqrt{N}} \left( \begin{array}{ccccc} 1 & 1 & \cdots & 1 & 1 \\  e^{2i\pi/N} &  e^{4i\pi/N} & \cdots & e^{2i\pi(N-1)/N}   & 1 \\  e^{4i\pi/N} &  e^{8i\pi/N} & \cdots & e^{4i\pi (N-1)/N}  & 1 \\ \vdots & \vdots & & \vdots & \vdots  \\ e^{2i\pi(N-2)/N} &  e^{4i\pi(N-2)/N} & \cdots &  e^{2i\pi(N-2)(N-1)/N}  &  1  \\   e^{2i\pi (N-1)/N} &  e^{4i\pi(N-1)/N} & \cdots & e^{2i\pi(N-1)^2/N}   & 1 \end{array} \right)=\frac{1}{\sqrt{N}}\, e^{2i\pi  m(l-1)/N}
\eea
with $(U^\dagger)_{lm}=\frac{1}{\sqrt{N}}\, e^{-2i\pi l(m-1)/N}$. Therefore, we get
\bea
(\Lambda^{-1})_{lm}=\frac{1}{N}\sum_k \kappa_k\, e^{-2i \pi(k-1)(l-m)/N}. 
\eea
We note that the number of two-fold degeneracies for $\kappa_k$ is given by $[\frac{N-1}{2}]$, which is determined by the number of a pair of integers, $n_1$ and $n_2$, satisfying $N=n_1+n_2$. (Here, $[A] $ means the largest integer less than or equal to  $A$.) For even $N=2n$, we get $[\frac{N-1}{2}]=n-1$ two-fold degeneracies, such as $\kappa_1$, $\kappa_{n+1}$, $\kappa_2=\kappa_{2n}$, $\kappa_3=\kappa_{2n-1}$, $\cdots$, $\kappa_{n-1}=\kappa_{n+3}$, $\kappa_{n}=\kappa_{n+2}$. 
On the other hand, for odd $N=2n+1$, we get $[\frac{N-1}{2}]=n$ two-fold degeneracies, such as $\kappa_1$, $\kappa_2=\kappa_N$, $\kappa_3=\kappa_{N-1}$, $\cdots$, $\kappa_n=\kappa_{n+3}$, $\kappa_{n+1}=\kappa_{n+2}$. 

For instance, for $N=3$, we get one two-fold degeneracy, such as $\kappa_1=1/(\lambda +2\lambda')$ and $\kappa_2=\kappa_3=1/(\lambda-\lambda')$.
For $N=4$, we get again one two-fold degeneracy, such as 
$\kappa_1=1/(\lambda+2\lambda')$, $\kappa_2=\kappa_4=1/\lambda$ and $\kappa_3=1/(\lambda-2\lambda')$.  For $N=5$, we get two two-fold degeneracies, such as $\kappa_1=1/(\lambda+2\lambda')$, $\kappa_2=\kappa_5=1/(\lambda+2\lambda'\cos(2\pi/5))$, $\kappa_3=\kappa_4=1/(\lambda-2\lambda' \cos(\pi/5))$.

As a consequence, we obtain
\bea
(\Lambda^{-1} M^2)_l&=& \frac{1}{N}\sum_{k,m} \kappa_k\, e^{-2i \pi (k-1)(l-m)/N} \, m^2_{\chi,m} \nonumber \\
&=&\frac{m^2_\chi}{N}\sum_{k,m} \kappa_k\, e^{-2i \pi (k-1)(l-m)/N}+  \frac{1}{N}\sum_{k,m} \kappa_k\, e^{-2i \pi (k-1)(l-m)/N} \,m^2_m.
\eea
Thus, we can show that the effective potential (\ref{effpot}) becomes constant,
\bea
V_{1,{\rm eff}}&=& -\frac{1}{4} (M^2)_l (\Lambda^{-1} M^2)_l \nonumber \\
&=&  -\frac{m^4_\chi}{4N}   \sum_{k,l,m} \kappa_k\, e^{-2i \pi (k-1)(l-m)/N}-\frac{m^2_\chi}{4N} \sum_{k,l,m} \kappa_k\, e^{-2i \pi (k-1)(l-m)/N} \,m^2_m \nonumber \\
&&-\frac{1}{4N} \sum_{k,l,m} \kappa_k\, e^{-2i \pi (k-1)(l-m)/N} \,m^2_m m^2_l \nonumber\\
&=&{\rm constant}.
\eea

If $\langle \chi_a\rangle=0$, for $a=N-M+1, N-M+2,\cdots, N$, the corresponding effective potential (\ref{effpot2}) in Configuration II becomes
\bea
V_{1,{\rm eff}}= \frac{1}{4} ((M^2)^T\Lambda^{-1})_a \Lambda_{ab} (\Lambda^{-1} M^2)_b+{\rm constant} 
\eea

\medskip

\clearpage
\bibliographystyle{apsrev4-1}

\begin{thebibliography}{999}

\bibitem{Planck}
Y.~Akrami \textit{et al.} [Planck],
Astron. Astrophys. \textbf{641} (2020), A10
doi:10.1051/0004-6361/201833887
[arXiv:1807.06211 [astro-ph.CO]].




\bibitem{keck}
P.~A.~R.~Ade \textit{et al.} [BICEP and Keck],
Phys. Rev. Lett. \textbf{127} (2021) no.15, 151301
doi:10.1103/PhysRevLett.127.151301
[arXiv:2110.00483 [astro-ph.CO]].




\bibitem{ACT}
T.~Louis \textit{et al.} [ACT],
[arXiv:2503.14452 [astro-ph.CO]].


\bibitem{Han:2025cwk}
J.~Han, H.~M.~Lee and J.~H.~Song,
[arXiv:2506.21189 [hep-ph]].




\bibitem{hybrid}
A.~D.~Linde,
Phys. Rev. D \textbf{49} (1994), 748-754
doi:10.1103/PhysRevD.49.748
[arXiv:astro-ph/9307002 [astro-ph]].



\bibitem{natural}
K.~Freese, J.~A.~Frieman and A.~V.~Olinto,
Phys. Rev. Lett. \textbf{65} (1990), 3233-3236
doi:10.1103/PhysRevLett.65.3233






\bibitem{multiaxion1}
J.~E.~Kim, H.~P.~Nilles and M.~Peloso,
JCAP \textbf{01} (2005), 005
doi:10.1088/1475-7516/2005/01/005
[arXiv:hep-ph/0409138 [hep-ph]].



\bibitem{multiaxion2}
A.~de la Fuente, P.~Saraswat and R.~Sundrum,
Phys. Rev. Lett. \textbf{114} (2015) no.15, 151303
doi:10.1103/PhysRevLett.114.151303
[arXiv:1412.3457 [hep-th]].



\bibitem{z2}
K.~Deshpande, S.~Kumar and R.~Sundrum,
JHEP \textbf{21} (2020), 147
doi:10.1007/JHEP07(2021)147
[arXiv:2101.06275 [hep-ph]].



\bibitem{Lee:2022fkd}
H.~M.~Lee and A.~G.~Menkara,
Phys. Lett. B \textbf{834} (2022), 137483
doi:10.1016/j.physletb.2022.137483
[arXiv:2206.05523 [hep-ph]].


\bibitem{Lee:2023dcy}
H.~M.~Lee and A.~G.~Menkara,
Phys. Rev. D \textbf{107} (2023) no.11, 115019
doi:10.1103/PhysRevD.107.115019
[arXiv:2304.08686 [hep-ph]].




\bibitem{qgravity}
M.~Kamionkowski and J.~March-Russell,
Phys. Lett. B \textbf{282} (1992), 137-141
doi:10.1016/0370-2693(92)90492-M
[arXiv:hep-th/9202003 [hep-th]];
S.~M.~Barr and D.~Seckel,
Phys. Rev. D \textbf{46} (1992), 539-549
doi:10.1103/PhysRevD.46.539



\bibitem{relaxion}
P.~W.~Graham, D.~E.~Kaplan and S.~Rajendran,
Phys. Rev. Lett. \textbf{115} (2015) no.22, 221801
doi:10.1103/PhysRevLett.115.221801
[arXiv:1504.07551 [hep-ph]].





\bibitem{twin}
  Z.~Chacko, H.~S.~Goh and R.~Harnik,
  Phys.\ Rev.\ Lett.\  {\bf 96} (2006) 231802
  doi:10.1103/PhysRevLett.96.231802
  [hep-ph/0506256].




\bibitem{hook}
  A.~Hook,
  Phys.\ Rev.\ Lett.\  {\bf 120} (2018) no.26,  261802
  doi:10.1103/PhysRevLett.120.261802
  [arXiv:1802.10093 [hep-ph]].


\bibitem{discrete}
L.~M.~Krauss and F.~Wilczek,
Phys. Rev. Lett. \textbf{62} (1989), 1221
doi:10.1103/PhysRevLett.62.1221;
L.~E.~Ibanez and G.~G.~Ross,
Phys. Lett. B \textbf{260} (1991), 291-295
doi:10.1016/0370-2693(91)91614-2;
T.~Banks and M.~Dine,
Phys. Rev. D \textbf{45} (1992), 1424-1427
doi:10.1103/PhysRevD.45.1424
[arXiv:hep-th/9109045 [hep-th]].


\bibitem{clockwork}
G.~F.~Giudice and M.~McCullough,
JHEP \textbf{02} (2017), 036
doi:10.1007/JHEP02(2017)036
[arXiv:1610.07962 [hep-ph]];
H.~M.~Lee,
Phys. Lett. B \textbf{778} (2018), 79-87
doi:10.1016/j.physletb.2018.01.010
[arXiv:1708.03564 [hep-ph]].


\bibitem{Espinosa:2017aew}
J.~R.~Espinosa and T.~Konstandin,
Phys. Rev. D \textbf{97} (2018) no.5, 056020
doi:10.1103/PhysRevD.97.056020
[arXiv:1712.08068 [hep-ph]].


\bibitem{Garcia-Bellido:2001dqy}
J.~Garcia-Bellido and E.~Ruiz Morales,
Phys. Lett. B \textbf{536} (2002), 193-202
doi:10.1016/S0370-2693(02)01820-8
[arXiv:hep-ph/0109230 [hep-ph]].




\bibitem{maxtemp}
E.~W.~Kolb, D.~J.~H.~Chung and A.~Riotto,
AIP Conf. Proc. \textbf{484} (1999) no.1, 91-105
doi:10.1063/1.59655
[arXiv:hep-ph/9810361 [hep-ph]].



\bibitem{reheating}
S.~M.~Choi and H.~M.~Lee,
Eur. Phys. J. C \textbf{76} (2016) no.6, 303
doi:10.1140/epjc/s10052-016-4150-5
[arXiv:1601.05979 [hep-ph]].


\bibitem{HiggsR2}
S.~Aoki, H.~M.~Lee, A.~G.~Menkara and K.~Yamashita,
JHEP \textbf{05} (2022), 121
doi:10.1007/JHEP05(2022)121
[arXiv:2202.13063 [hep-ph]].





\end{thebibliography}

\end{document}